\documentclass[a4paper,11pt]{article}
\pdfoutput=1 
\usepackage{jheppub} 
\usepackage{tikz}
\usepackage{slashed}
\usepackage{ulem}
\usepackage{subfigure}
\usepackage{comment}
\usepackage{yhmath}
\usepackage[all]{xy}
\usepackage{multirow}
\usepackage{float}
\usepackage{mathtools}
\usepackage{xfrac}
\usepackage{wrapfig}
\usepackage{dsfont}
\usepackage{makecell}

\numberwithin{equation}{section}

\newcommand{\beq}{\begin{equation}}
\newcommand{\eeq}[1]{\label{#1}\end{equation}}

\newcommand{\SSM}{{\mathcal{SM}}}

\newcommand{\SM}{\mathcal{M}^{(4|4)} }
\def\a{\alpha}

\def\b{\beta}

\def\d{\delta}

\def\g{\gamma}

\def\r{\rho}
\def\s{\sigma}

\def\S{\Sigma}

\newcommand{\be}{\begin{equation}}
\newcommand{\ee}{\end{equation}}
\newcommand{\bea}{\begin{eqnarray}}
\newcommand{\eea}{\end{eqnarray}}

\newcommand{\ba}{\begin{array}}
\newcommand{\ea}{\end{array}}

\def\double #1{#1{\hbox{\kern-2pt $#1$}}}

\newcommand{\bsubeq}{\begin{subequations}}
 
\newcommand{\esubeq}{\end{subequations}}

\begin{document}

\begin{titlepage}

\begin{flushright}
\par\end{flushright}
\vskip 0.5cm
\begin{center}
\textbf{\LARGE \bf 
Super-Higher-Form Symmetries }\\
\vskip 5mm
\end{center}


\begin{center}

\large 
{\bf 
 \large {\bf P.~A.~Grassi}$^{~a,b,c,}$\footnote{pietro.grassi@uniupo.it} 
\bf  
 \large {\bf S.~Penati}$^{~d,}$\footnote{silvia.penati@unimib.it} 

 \vskip .5cm 
\small

 {$^{(a)}$
 \it DiSIT,} 
 {\it Universit\`a del Piemonte Orientale,} 
 {\it viale T.~Michel, 11, 15121 Alessandria, Italy}
 }
 \small
 
 {$^{(b)}$
 \it Theoretical Physics Department, CERN, 1211 Geneva 23, Switzerland} 
\small
 
 {$^{(c)}$
 \it INFN,} 
 {\it Sez. Torino,} 
 {\it via P. Giuria, 1, 10125, Torino, Italy}
\small

{$^{(d)}$
 {\it Dipartimento di Fisica, Universit\`a degli Studi di Milano--Bicocca and INFN, Sezione di Milano--Bicocca, Piazza della Scienza 3, 20126 Milano, Italy}
 }

\end{center}
\vspace{1cm}

\vskip  0.2cm
\begin{abstract}

We generalize the study of higher-form-symmetries to theories with supersymmetry. Using a supergeometry formulation, we find that ordinary higher-form-symmetries nicely combine with supersymmetry to give rise to a much larger spectrum of topological conserved (super)currents. These can be classified as a supersymmetric version of Chern-Weil symmetries, and a brand new set of geometric-Chern-Weil symmetries whose generators are constructed using invariant differential forms in supermanifolds. For N=1 super-Maxwell theory in various dimensions, we build the topological operators  generating these super-higher-form symmetries and  construct 
defects carrying non-trivial charges. Notably, the charge is proportional to the super-linking number between the super-hypersurface supporting the symmetry generator and the one supporting the defect.

\end{abstract}
\vfill{}
\vspace{1.5cm}
\end{titlepage}
\newpage\setcounter{footnote}{0}

\tableofcontents
\vspace{1cm}

\section{Introduction}

In recent years, higher-form global symmetries have attracted increasing interest due to their connection with sectors of higher dimensional operators (defects) and the role played in further constraining the dynamics of quantum systems, thanks to new Ward identities, anomalies, breakings or gaugings. This has led to a new paradigm in how we think about symmetries in QFT, where topological and non-topological defects play an instrumental role.   

Generalized symmetries have been classified as invertible \cite{Gaiotto:2014kfa} and non-invertible symmetries, the latter having a long history in two-dimensional systems \cite{Frohlich:2004ef}, then generalized to higher dimensional ones only recently (see for instance \cite{Schafer-Nameki:2023jdn,Shao:2023gho} for a review on recent developments). Our focus will be on invertible higher-form global symmetries. A nice introduction to this topic and an exhaustive list of references can be found in \cite{Brennan:2023mmt,Gomes:2023ahz,Luo:2023ive}. 

Higher-form symmetries can be conveniently framed into a geometrical scheme based on differential geometry, differential form formalism, and hypersurface geometry. One of the most important observations is that the field strength $F^{(p)}= dA^{(p-1)}$ of any local $(p-1)$-form gauge potential $A^{(p-1)}$ is a conserved current, thanks to the Bianchi identity. In $n$ dimensions, the topological charge operator $Q(\Sigma_{p}) = \exp(i\int_{\Sigma_{p}} F^{(p)})$ generates a $(n-p-1)$-form symmetry under which  $(n-p-1)$-dimensional defects are charged. More generally, given a set of closed field strengths $F^{(p)}$ for different values of $p$, higher-form currents can be constructed as products $F^{(p_1)} \wedge \dots \wedge F^{(p_k)}$, leading to newly-explored Chern-Weil symmetries \cite{Heidenreich:2020pkc}. Since the seminal paper \cite{Gaiotto:2014kfa}, these symmetries and their properties - anomalies, breaking, or gauging - have been extensively studied in gauge theories with or without matter, in various dimensions (see references in \cite{Brennan:2023mmt,Gomes:2023ahz,Luo:2023ive}). A deeply studied case is the $2n$-dimensional axion-electrodynamics \cite{Hidaka:2020iaz,Hidaka:2021kkf,Nakajima:2022feg}  and its non-abelian generalization \cite{Brauner:2020rtz,Anber:2024gis}. 

In supersymmetric theories, topological higher-form symmetries combine with supersymmetry to give rise to a new unexplored net of higher-form conservation laws, whose effects on the quantum dynamics of the system need to be investigated, along with the construction of non-local operators charged under them. In \cite{Cremonini:2020zyn} we gave a first grasp of the general picture. In this paper we will pursue a more systematic investigation. 

The possibility to define a supersymmetric theory on a supermanifold\footnote{For an introduction to supersymmetric theories in supermanifolds, see \cite{ 
Fre:2017bwd,Castellani:2023tip}.}, or more simply on a superspace, which is the local version of a supermanifold, allows for accomplishing this program in a very efficient way. In fact, several ingredients discussed in the aforementioned literature can be easily translated into a supersymmetric framework. First of all, differential calculus in supermanifolds \cite{BL1, BL2, Manin,V1,V2,V3,V4,Belopolsky:1997bg,Belopolsky:1997jz,Voro}, leads to the notion of superdifferential forms $J^{(p|q)}$ that locally can be expressed in terms of fundamental $(1|0)$-forms $dx$’s and $d\theta$’s related to the even and odd coordinates of the superspace, with the coefficients of the expansion being ordinary superfields. In a $(n|m)$-dimensional supermanifold, they are featured by a form number $0\leq  p \leq n$ and a so-called picture number $0\leq q \leq m$. Superdifferential forms satisfying the closure condition $d J^{(p|q)}=0$ are the natural candidates for defining  {\em super-higher-form currents}, that is, the generalization of higher-form conserved currents to a supersymmetric set-up. The general theory of integration on supermanifolds \cite{Witten:2012bg} allows then to define the corresponding generators as integrals of the currents on $(p|q)$-dimensional super-hypersurfaces $\Sigma_{(p|q)}$ embedded into the supermanifold,
$Q(\Sigma_{(p|q)}) = \int_{\Sigma_{(p|q)}} \!\! J^{(p|q)}$.
Thanks to the closure of the super differential form, these are topological operators that generate a $(n-p-1|m-q)$-higher-form symmetry. Charged objects can be constructed in general, which are operators localized on a $(n-p-1|m-q)$—hypersurface.

A first formal discussion of super-higher-form currents was given in \cite{Cremonini:2020zyn}, where for generic tensorial superform currents $J^{(p|0)}$ satisfying $dJ^{(p|0)}=0$, the corresponding topological supercharge was identified and the non-local operators in supermanifold charged under this super-higher-form symmetry were constructed. The interesting result is that the expansion of $J^{(p|0)}$ on a supervielbein basis in the supermanifold provides a new set of $(p-1)$ conserved supercurrents, which in turn give rise to current supermultiplets, once expanded in powers of the spinorial coordinates. Fermionic higher-form symmetries have been recently investigated in \cite{Chang:2022hud,Wang:2023iqt,Ambrosino:2024ggh}, not necessarily in a supersymmetric context. 

In this paper, we specialize this construction to abelian gauge theories in various dimensions and with different amounts of supersymmetry. We focus not only on {\em superform} $J^{(p|0)}$  currents but also on {\em integral form} $J^{(p|m)}$ currents - where $m$ is the odd dimension of the supermanifold. The construction of more general $J^{(p|q)}$ symmetries and the discussion of their physical meaning is left for future investigation. 
We present the geometry of the topological charge operators and the set of objects charged under them. We postpone the study of their physical implications, associated anomalies, gauging, and breaking - also in the context of supergravity - to a subsequent paper \cite{grassi}.

Abelian gauge theories in supermanifolds are described by a $(1|0)$-superform gauge potential $A$, whose superfield strength $F = dA$ is a $(2|0)$-superform satisfying $dF=0$. In principle, this superform contains too many degrees of freedom. An important ingredient is the choice of the constraints that reduce the number of its independent components. In any case, the bottom line is that Bianchi identities $d  F=0 $, together with the equations of motion $d \star F=0 $, are always the crucial key to establishing the content of the physical multiplets, their independent degrees of freedom, and their representations. In several cases, Bianchi identities also imply the equations of motion. This is true for both rigid supersymmetry (see \cite{Harnad:1985bc} and the textbooks \cite{Wess:1992cp,Gates:1983nr}) and supergravity (see for example \cite{Castellani:1991et}). 

The superfield strength $F $ naturally gives a well-defined conserved super-higher-form current. It is the lowest representative of topological superform currents of type $J^{(p|0)}$. We study the corresponding generator that gives rise to a $(n-3|m)$-super-higher-form symmetry. We also identify the objects charged under this symmetry as 't Hooft-like supermonopoles supported on a singular $(n-3|m)$-dimensional super-hypersurface. 
In addition, depending on the bosonic dimensions of the supermanifold, higher-superform currents can be defined by taking products of $k$ $F$ factors, $J^{(2k|0)} = F \wedge F \wedge \dots F$. They give rise to what we call super-Chern-Weil symmetries. Suitable multi-supermonopole configurations can be constructed, which carry a $J^{(2k|0)}$ charge.

Moving to the class of integral form currents, the lowest dimensional representative is given by the on-shell closed current $J^{(n-2|m)} = \star F$, generating a $(1|0)$-integral form symmetry. This is the generalization to supermanifolds of the Noether-like current induced by the Maxwell equations of motion. Objects charged under this symmetry are supersymmetric Wilson loops supported on $(1|0)$-supercycles \cite{Cremonini:2020mrk}.

In a supersymmetric context the novelty is that, beyond the brand new set of supercurrents which originate from the reduction to components of superform and integral form currents constructed from $F$ and $\star F$,  another set of conserved currents can be determined, owing to the existence of supersymmetric invariant expressions in the cohomology of supermanifolds. These invariants are constructed in terms of supervielbeins, and their invariance is related to the Fiersz identities, which are specific to any superspace. We can multiply these invariants by topological super-higher-form currents of the form $F \wedge \dots F$ to give rise to the generators of new symmetries. We dub these new symmetries super-geometric-Chern Weil symmetries. We explore all these possibilities in the examples of super-Maxwell theories we consider. Notably, we prove that supermonopoles are also charged under the action of these new symmetry generators.

\vskip 10pt

The rest of the paper is organized as follows in section \ref{sec:section1} we briefly recall the construction of higher-form currents in ordinary gauge theories, focusing in particular on $n$-dimensional Maxwell theory. We work in a geometric set-up, fixing notions and notations in a language that is easy to generalize to supermanifolds. We discuss the coupling of the higher-form currents to background gauge fields as an efficient prescription for determining the action of symmetry generators on charged operators. Section \ref{sec:supersymmetries} is the core of the paper, where the construction of super-higher-form symmetries in supermanifolds is discussed in general, focusing on $J^{(p|0)}$ superform and $J^{(p|m)}$ integral form currents. In section \ref{sec:superMaxwell}, this construction is applied to specific examples of $N=1$ supersymmetric Maxwell theory, precisely in three, four, six, and ten dimensions. In some of these cases, we also rewrite in supergeometry language the explicit expression of the generators of ordinary supersymmetry in order to facilitate a comparison with the structure of super-higher-form generators. We discuss in detail the construction of operators charged under these new symmetries and determine their charge as given by the super-linking number, the generalization of the linking number to supermanifolds \cite{Cremonini:2020zyn}. We complete the spectrum of applications by discussing in section \ref{sec:CS} the interesting case of the abelian $N=1$ super-Chern-Simons theory in three dimensions.
Finally, we draw our conclusions in section \ref{sec:conclusions} and highlight possible directions for further investigation. For readers not acquainted with the language of supergeometry, we collect a little information in appendix 
\ref{app:superforms} and review the definition of (super)linking number in appendix \ref{app:linking}. Finally, in appendix \ref{sect:CWcurrents}, we report details on the construction of operators charged under super-Chern-Weil and super-geometric-Chern-Weil symmetries for Maxwell theory in $D=(6|8)$ supermanifold. 

\section{Ordinary Higher-Form Symmetries}
\label{sec:section1}

We begin by briefly recalling the construction of higher-form global symmetry generators and the corresponding charged operators. We rephrase known material \cite{Gaiotto:2014kfa, Heidenreich:2020pkc, Brauner:2020rtz,Nakajima:2022feg} (see also reviews  \cite{Brennan:2023mmt,Gomes:2023ahz,Luo:2023ive}) in a geometrical language, as this is suitable for the generalization to the supersymmetric case that we develop in section \ref{sec:supersymmetries}. 

In this paper, we focus only on abelian generalized global symmetries. They can be either Noether-like when conservation follows from some equations of motion, or topological when conservation comes from constraints, typically Bianchi identities.

Given a classical field theory defined in a $n$-dimensional oriented manifold $\mathcal{M}$, a  $p$-form symmetry is generated by a conserved $(n-p-1)$-form current $J^{(n-p-1)} = \star J^{(p+1)}$. The corresponding conserved charge is 
\begin{equation}
Q(\Sigma_{n-p-1})  = \int_{\Sigma_{n-p-1}} J^{(n-p-1)} \, ,
\end{equation}
where $\Sigma_{n-p-1}$ is a $(n-p-1)$-dimensional compact oriented submanifold in $\mathcal{M}$. 

It is more convenient to rewrite the $Q$ integral  by introducing the Poincar\'e dual $\mathbb{Y}_{\Sigma_{n-p-1}}^{(p+1)}$ which parametrizes the embedding of $\Sigma_{n-p-1}$ inside ${\mathcal M}$. We recall that if the cycle is defined by a set of equations $\phi_i(x^a) = 0$, $i = 1, \dots, p+1$ in the manifold described by $x^a$ coordinates, the Poincar\'e dual is given by
\begin{equation}
\label{eq:PCO}
 \mathbb{Y}_{\Sigma_{(n-p-1)}}^{(p+1)} = \prod_{i=1}^{p+1} \delta(\phi_i) d \phi_i \, .
\end{equation}

For a closed cycle, $\mathbb{Y}_{\Sigma_{n-p-1}}^{(p+1)}$ is closed but not exact, and a change in the cycle corresponds to a $d$-exact term. Precisely, 
\begin{eqnarray}
\label{WLB}
d \mathbb{Y}_{\Sigma_{n-p-1}}^{(p+1)} = 0\,, ~~~~~~~
\mathbb{Y}_{\Sigma_{n-p-1}}^{(p+1)} \neq d Z^{(p)}\,, ~~~~~~~
\delta \mathbb{Y}_{\Sigma_{n-p-1}}^{(p+1)} = d \Lambda^{(p)} \, .
\end{eqnarray}

Thanks to these properties and the current conservation law, the charge can be rewritten as an integral on the entire manifold,
\begin{equation}
Q(\Sigma_{n-p-1})  = \int_{\mathcal M} J^{(n-p-1)} \wedge \mathbb{Y}_{\Sigma_{n-p-1}}^{(p+1)} \, .
\end{equation}
Moreover, it is manifestly conserved and topological (it does not depend on the choice of $\Sigma_{n-p-1}$).
It generates $p$-form symmetry transformations on the fields, parametrized by a closed $p$-form  parameter. 

Operators charged under this symmetry are $p$-dimensional cycle operators ($p$-dimensional defects) of the form
\begin{equation}
\label{eq:Wq}
    W(\Sigma_p) = e^{iq \Gamma} \, , \qquad \Gamma = \int_{\Sigma_p} A^{(p)}_* = \int_{\mathcal M} A^{(p)} \wedge \mathbb{Y}_{\Sigma_p}^{(n-p)}\, , 
\end{equation}
for a given $p$-form connection $A^{(p)}$, supported on a compact oriented $p$-dimensional submanifold $\Sigma_p$. Here $\mathbb{Y}_{\Sigma_p}^{(n-p)}$ is the Poincar\'e dual of the $p$-cycle $\S_p$. In writing $\Gamma$, we made a distinction  between $A^{(p)}$ and $A^{(p)}_*$, which is the pullback of 
$A^{(p)}$ on the $p$-cycle. To avoid cluttering, in the rest of the paper we will neglect such a distinction. 

An alternative way of writing $\Gamma$ is obtained by introducing a $(p+1)$-dimensional hypersurface $\Omega_{p+1} \subset {\mathcal M}$ whose boundary is $\Sigma_p$. Due to the presence of a boundary, the corresponding  Poincar\'e dual 
$\Theta_{\Omega_{p+1}}^{(n-p-1)}$ is not closed, rather it satisfies $d\Theta_{\Omega_{p+1}}^{(n-p-1)} = \mathbb{Y}_{\Sigma_p}^{(n-p)}$\footnote{In our conventions, $\mathbb{Y}$ always denotes a Poincar\'e dual with compact support ($d \mathbb{Y} =0$), whereas 
$\Theta$ is used for Poincar\'e dual forms of submanifolds with boundaries, therefore not closed.}. Now, defining $F^{(p+1)} \equiv dA^{(p)}$, we can write
\begin{equation}
\label{eq:Wq2}
\Gamma = \int_{\Omega_{p+1}} F^{(p+1)} = \int_{\mathcal M} F^{(p+1)} \wedge \Theta_{\Omega_{p+1}}^{(n-p-1)} \, .
\end{equation}
The equivalence of this expression with \eqref{eq:Wq} can be straightforwardly proved by integrating by parts. 

In canonical formalism, the action of the symmetry generator on $W$ is given by (for a derivation of this identity see for instance \cite{Gomes:2023ahz})
\begin{equation}
\label{eq:QonW}
    e^{i\a Q(\Sigma_{n-p-1}) } \, W(\Sigma_p) \, e^{-i\a Q(\Sigma_{n-p-1}) }  = e^{i\a \, q \, {\rm Link}(\S_{n-p-1}, \S_p)} \, W(\Sigma_p)\, ,   
\end{equation}
where ${\rm Link}(\S_{n-p-1}, \S_p)$ is the linking number between the two cycles. It is defined as \cite{botttu} (see appendix \ref{app:linking} for details)
\begin{equation}
\label{eq:link}
{\rm Link}(\S_{n-p-1}, \S_p) \equiv 
\int_{\mathcal M}   \mathbb{Y}_{\Sigma_{n-p-1}}^{(p+1)}
\wedge \Theta_{\Omega_{p+1}}^{(n-p-1)} = (-1)^{p+1} \int_{\mathcal M}   \Theta_{\Omega_{n-p}}^{(p)} 
\wedge \mathbb{Y}_{\Sigma_p}^{(n-p)} 
\, ,   
\end{equation}
where $\partial \Omega_{p+1} = \Sigma_p$ , whereas $\Omega_{n-p}$ is the $(n-p)$-dimensional hypersurface whose boundary is $\S_{n-p-1}$. Again, due to the effect of the boundary, its Poincar\'e dual is not closed, rather it satisfies  $d \Theta_{\Omega_{n-p}}^{(p)} = \mathbb{Y}_{\Sigma_{n-p-1}}^{(p+1)}$. The two alternative ways of defining the Link simply differ by an integration-by-parts. The linking number is non-vanishing only if the dimensions of two hypersurfaces sum up to $(n-1)$ and the supports of the two integrated Poincar\'e duals in \eqref{eq:link}  have a non-trivial intersection, equivalently if the two surfaces $\Sigma_{n-p-1}$ and $\Sigma_p$ are disjoint.

The transformation in \eqref{eq:QonW} corresponds to a translation $A^{(p)} \rightarrow  A^{(p)} + i \a  \, \Theta_{\Omega_{n-p}}^{(p)}$ of the connection appearing in $W$. 

Identity \eqref{eq:QonW}can be alternatively recovered using the path integral formulation. Since this approach turns out to be more convenient for treating Noether-like and topological symmetries on the same foot and for generalizing the construction to supersymmetric theories, we review it in the next subsection for the special case of Maxwell theory.

\subsection{Maxwell theory in $n$ dimensions}
\label{sec:maxwell}

As an illustrative example, we review the construction of higher-form symmetries in pure Maxwell theory defined on a $n$-dimensional manifold ${\mathcal M}$. 

From the usual action \begin{equation}
\label{eq:maction}
    S  = - \frac12 \int_{\mathcal M} F \wedge \star F \, ,
\end{equation}
with $F= dA$, the equations of motion 
$d \star F= 0$ and the Bianchi identities $d F = 0$ provide two conservation laws. The corresponding conserved charges 
\begin{eqnarray}
\label{eq:Qe}
    && Q_e (\Sigma_{n-2}^e)= \int_{\Sigma_{n-2}^e} \!\! \! \!  \star F = \int_{\mathcal M} \star F \wedge \mathbb{Y}_{\Sigma_{n-2}^e}^{(2)} \, , \\ 
    \label{eq:Qm}
    && Q_m (\Sigma_2) = \int_{\Sigma_2}  F = \int_{\mathcal M} F \wedge \mathbb{Y}_{\Sigma_2}^{(n-2)} 
\end{eqnarray}
generate two U(1) higher-form symmetries, a Noether-like 1-form symmetry and a topological $(n-3)$-form one, respectively.
Thanks to the introduction of the Poincar\'e dual operators, the charges have been defined as integrals on the whole manifold and are manifestly independent of the choice of the cycles. 

For manifolds of dimension $n > 2k$, 
one can build more general higher-form conserved currents by taking wedge products of $F$ 
\begin{eqnarray}
    \label{CWA}
    J^{(2k)} = \frac{1}{k!} \, \underbrace{F \wedge F \wedge \dots F}_k \, .
\end{eqnarray}
They correspond to conserved topological charges 
\begin{eqnarray}
    \label{CWB}
    Q^{(k)} (\Sigma_{2k})=\int_{\mathcal M} J^{(2k)} \wedge \mathbb{Y}^{(n-2k)}_{\Sigma_{2k}} \, ,
\end{eqnarray}
generating $(n-2k-1)$-form U(1) symmetries. These are named Chern-Weil (CW) currents and have been discussed in \cite{Heidenreich:2020pkc, Brauner:2020rtz, Nakajima:2022feg}. The lowest representative $Q^{(1)}$ coincides with the magnetic charge \eqref{eq:Qm}. In principle, one can consider using a factor $\star F$. The only possibility is to take the lagrangian $F \wedge \star F$. This current saturates the dimensions of the manifold and gives rise to a Chern-Weil $(-1)$-form symmetry \cite{Cordova:2019jnf,McNamara:2020uza,Aloni:2024jpb,
Brennan:2024tlw,Najjar:2024vmm}. We will not further consider this kind of symmetry. 

\vskip 10pt

In order to study the action of $Q_e$ and $Q^{(k)}$, $k \geq 1$ on charged operators, it is convenient to minimally couple the corresponding conserved currents to a set of background gauge fields.
This can be done in different ways, and depending on the particular choice of the couplings, anomalies may arise for different currents.\footnote{Different couplings differ by redefinitions of the gauge fields and/or the addition of extra pure background terms.} We choose to consider the following total action \cite{Nakajima:2022feg} 
\begin{eqnarray}
    \label{GcwB0}
    S + S_{min} &=&  - \frac12 \int_{{\mathcal M}} F\wedge \star F  + 
  \int_{{\mathcal M}} \left(    \star F \wedge B^{(2)} + 
     \sum_k J^{(2k)}\wedge B^{(n-2k)}_{CW} 
    \right) \\
   & & \hspace{-1cm} = \int_{{\mathcal M}} \left( - \frac12 (F - B^{(2)}) \wedge \star (F - B^{(2)}) +  F \wedge B_{CW}^{(n-2)} + 
    \sum_{k>1} J^{(2k)}\wedge B^{(n-2k)}_{CW} + \frac12  B^{(2)} \wedge \star B^{(2)}) 
    \right) \, , \nonumber 
\end{eqnarray}
with $J^{(2k)}$ given in \eqref{CWA} for $2k<n$. The pure background term $\frac12 B^{(2)} \wedge \star B^{(2)}$ can be neglected, as it can be absorbed into the path integral normalization. 

To be definite, we fix $n=6$. In this simple but sufficiently instructing case, we have three higher-form currents, the Noether-like current $J = \star F$, and the two topological currents $J^{(2)} = F$ and $J^{(4)} = \frac12 F \wedge F$.
They couple to background gauge fields $B^{(2)}, B^{(4)}_{CW}$ and $B^{(2)}_{CW}$, respectively. 

The action in \eqref{GcwB0} is not invariant under gauge transformations generated by these currents. However, gauge invariance can be restored - up to anomalies - by imposing non-trivial gauge transformations on the background fields. Under the  following transformation laws
\begin{eqnarray}
\label{eq:transfs1}
    && F \rightarrow F + d\lambda^{(1)} 
    \, , \nonumber \\
    && B^{(2)} \rightarrow B^{(2)} + d \lambda^{(1)} \, ,\nonumber \\
    && B_{CW}^{(4)} \rightarrow B_{CW}^{(4)} + d\lambda^{(3)} - d\lambda^{(1)} \wedge B^{(2)}_{CW} \, ,\nonumber \\
    && B^{(2)}_{CW} \rightarrow B^{(2)}_{CW} + d \lambda_{CW}^{(1)} \, ,
\end{eqnarray}
the action is invariant up to the following 't Hooft anomaly 
\begin{eqnarray}
\label{eq:anomaly0}
S' + S_{min}' &=& S + S_{min} + \int_{\mathcal M}  \left( d \lambda^{(1)} \wedge B_{CW}^{(4)} - \frac12 d \lambda^{(1)} \wedge d \lambda^{(1)} \wedge B^{(2)}_{CW} \right)\, .
\end{eqnarray} 

It is important to stress that with our choice of gauge couplings and background gauge transformations, the only anomalous symmetry turns out to be the Noether-like one-form symmetry. The action is instead invariant under the topological U(1) three-form and one-form symmetries generated by $J^{(2)}$ and $J^{(4)}$ respectively, as long as the corresponding currents are conserved, that is, there are no monopoles/defects. 

Introducing a suitable gauge fixing term $S_{gf}$, up to a normalization factor, the path integral is defined as 
 \begin{eqnarray}
     \label{GcwC0}
      Z[B^{(2)}, B_{CW}^{(4)}, B^{(2)}_{CW}] &=& \int [\mathcal{D} A] \, e^{ i \, ( S +S_{min} + S_{gf})} \, .
 \end{eqnarray}
The action of the symmetry generators 
\begin{eqnarray}
\label{GcwD0}
U(\Sigma_4^{e}, \alpha_e) = e^{i \alpha_e Q_e} \, , \qquad \quad 
U(\Sigma_{2k}, \alpha_k) = e^{i \alpha_k Q^{(k)}}
\end{eqnarray}
on charged operators $W$ is obtained from the two-point function $\langle U W \rangle$, that is from \eqref{GcwC0} with the insertion of $U$ and $W$.

We first consider the $Q_e$ symmetry. Operators charged under \eqref{eq:Qe} 
are ordinary Wilson loops of the form
\begin{eqnarray}
    \label{GcwE}   W(\mathcal{C}) = e^{i q \int A \wedge \mathbb{Y}^{(5)}_{\mathcal{C}}} = e^{i q \int F \wedge \Theta_{\Omega_2}^{(4)}} \, ,
\end{eqnarray}
where $\mathcal{C}$ is the line describing the boundary of $\Omega_2$, and  $\mathbb{Y}^{(5)}_{\mathcal{C}}$ is its Poincaré dual form satisfying $\mathbb{Y}^{(5)}_{\mathcal{C}} = d \Theta^{(4)}_{\Omega_2}$.\footnote{If the contour ${\mathcal C}$ is described by the map $\tau \rightarrow x^a(\tau)$, it is convenient to identify $\tau$ with one coordinate of the manifold, such that a representative of $\mathbb{Y}_{\mathcal{C}}^{(5)}$ can be explicitly written as 
\begin{eqnarray}
\label{WLC}
\mathbb{Y}_{\mathcal{C}}^{(5)} =\prod_{a=1}^{5}  \delta(x^a - x^a(\tau)) \bigwedge_{a=1}^{5}(dx^a - \dot x^a d\tau)  \, .
\nonumber
\end{eqnarray}
.} Gauge invariance is ensured by the closeness of $\mathbb{Y}_{\mathcal{C}}^{(5)}$.
Instead, a deformation of the contour, which is equivalent to changing 
$\mathbb{Y}_{\mathcal{C}}^{(5)} \rightarrow \mathbb{Y}_{\mathcal{C}}^{(5)} + d \Lambda^{(4)}$, may lead to a non-trivial change in the Wilson loop,  
$\delta \Gamma = \int F \wedge   \Lambda^{(4)} $, 
unless $F =0$ 
on the surface connecting the loop and its deformation if the curve $\mathcal{C}$ has been deformed without encountering singularities. 

From (\ref{GcwD0}, \ref{eq:Qe}) it is easy to see that the insertion of the generator $U(\Sigma^e_{4}, \a_e) $  inside the path integral corresponds to turning on a background gauge field $B^{(2)} = \a_e \mathbb{Y}^{(2)}_{\Sigma^e_{4}}$. Analogously, the insertion of $W(\mathcal{C})$ corresponds to turning on $B_{CW}^{(4)} = q \Theta^{(4)}_{\Omega_2}$. Therefore, the two-point function is nothing but
\begin{equation}
    \label{GcwF}
    \Big\langle U(\Sigma^e_{4}, \alpha_e) \, W(\mathcal{C}) \Big\rangle =
    Z[\alpha_e \mathbb{Y}^{(2)}_{\Sigma^e_4}, q \Theta^{(4)}_{\Omega_2}, 0] \, .
\end{equation}
Now we gauge away $B^{(2)}$ by performing a gauge  transformation \eqref{eq:transfs1} with $d\lambda^{(1)} = - \a_e \mathbb{Y}^{(2)}_{\Sigma^e_4}$, thus obtaining $Z[0, q \Theta^{(4)}_{\Omega_2}, 0] = \langle W(\mathcal{C}) \rangle$. However, as mentioned above, this gauge transformation is anomalous. Taking into account the explicit expression \eqref{eq:anomaly0} for the anomaly, one eventually finds 
\begin{equation}
    \label{GcwF2}
    \Big\langle U(\Sigma^e_4, \alpha_e) \, W(\mathcal{C}) \Big\rangle  = e^{i \alpha_e q \int  \mathbb{Y}^{(2)}_{\Sigma^e_4} \wedge \Theta^{(4)}_{\Omega_2}} \, \Big\langle W(\mathcal{C})\Big\rangle \, .
\end{equation} 
We conclude that the Wilson loop is the operator charged under the Noether-like 1-form symmetry, with charge given by the linking number ${\rm Link}({\mathcal C}, \Sigma^e_4) = \int  \mathbb{Y}^{(2)}_{\Sigma^e_4}\wedge \Theta^{(4)}_{\Omega_2}$. Interestingly, the charge is entirely due to the 't Hooft anomaly. 

We now move to the 3-form symmetry generated by $J^{(2)} = F$. 
In six dimensions, the corresponding charged object is a monopole supported on a 't Hooft surface $\Sigma_3 \subset {\mathcal M}$, the boundary of a four-dimensional submanifold $\Omega_4$. The corresponding Poincar\'e duals satisfy $Y^{(3)}_{\Sigma_3} = d \Theta^{(2)}_{\Omega_4}$. 

In the presence of a monopole, the conservation law $dF=0$ is no longer valid everywhere. We split the field strength into a regular and a singular part, $F = F_{r} + F_{s}$, satisfying
\begin{eqnarray}
 d F_{r} = 0  \quad && \Rightarrow \quad F_{r} = d A_{r} \nonumber \\  
  d F_{s} = q \mathbb{Y}^{(3)}_{\Sigma_3} = q \, d \Theta_{\Omega_4}^{(2)} \quad && \Rightarrow \quad F_{s} = q \, \Theta_{\Omega_4}^{(2)} \, .
\end{eqnarray}
Here $q$ is the charge of the monopole. It satisfies the Dirac quantization condition 
$\oint_{S^2} B^{(2)}_{\mathcal{D}} = 2 \pi q$, 
with $S^2$ being a two-dimensional sphere embedded in the three-dimensional submanifold transversal to $\Sigma_3$.

Since we integrate only on $A_r$ in the path integral, the replacement  $F \to F_r+F_s$ in the action gives rise to the term $\int \star F_r \wedge F_s$, which in turn corresponds to turning on a background $B^{(2)}_{\mathcal{D}} = F_s = q \Theta_{\Omega_4}^{(2)}$.

Therefore, the expectation value of the monopole is given by
\begin{equation}
\label{eq:monopole}
\Big\langle \mathcal{D}(\Sigma_3) \Big\rangle = \int [ {\mathcal D} A_r] \, e^{i S[F_r + F_s]} = Z[q \Theta_{\Omega_4}^{(2)}, 0, 0] \, .
\end{equation}
On the other hand, the insertion of the generator $U(\Sigma_2, \a_1)$ in (\ref{GcwD0}, \ref{eq:Qm}) corresponds to turning on a background field $B_{CW}^{(4)} = \a_1 \mathbb{Y}^{(4)}_{\Sigma_2}$.
Therefore, the two-point function is the path integral 
\begin{equation}
\Big\langle U(\Sigma_2, \a_1) \, \mathcal{D}(\Sigma_3) \Big\rangle = 
Z[ q \Theta^{(2)}_{\Omega_4}, \a_1 Y^{(4)}_{\Sigma_2},  0] \, . 
\end{equation}
We now perform a U(1) gauge transformation \eqref{eq:transfs1} with $d\lambda^{(3)} = - \a_1 Y^{(4)}_{\Sigma_2}$ to get rid of the background gauge field $B_{CW}^{(4)}$.  In the presence of the monopole, the action is no longer invariant; rather, we obtain 
\begin{equation}
S[F, B^{(2)}_{\mathcal{D}}, B_{CW}^{(4)}, 0]  = 
S[F, B^{(2)}_{\mathcal{D}}, 0, 0] - \int F_s \wedge d\lambda^{(3)} \, .
\end{equation}
Inserting the explicit expressions for $F_s$ and $d \lambda^{(3)}$, we end up with the following identity
\begin{eqnarray}
    \label{GcwE2}
    \Big\langle U(\Sigma_2, \a_1) \, \mathcal{D}(\Sigma_3) \Big\rangle = e^{i \a_1 q \int  \mathbb{Y}^{(4)}_{\Sigma_2}\wedge \Theta^{(2)}_{\Omega_4}} \, \Big\langle \mathcal{D}(\Sigma_3) \Big\rangle \, .
\end{eqnarray}
Again, the `t Hooft anomaly provides the operator's charge, which is proportional to the linking number 
between $\Sigma_2$ and $\Sigma_3$. 

Finally we consider the conserved current $J^{(4)}$ in \eqref{CWA}. 
Since it generates a 1-form symmetry, the corresponding charged operator is a one-dimensional line defect supported by $\tilde{\mathcal C}$, whose Poincaré dual satisfies $ \mathbb{Y}^{(5)}_{\tilde{\mathcal C}} = d  \Theta^{(4)}_{\tilde\Omega_2}$, with $\partial \tilde\Omega_2 = \tilde{\mathcal C}$. The line defect can be realized as the intersection of two monopole surfaces $\Sigma_3, \Sigma_3'$, which have one dimension in common \cite{Nakajima:2022feg}.
It acts as a localized source for $J^{(4)}$, precisely
\begin{equation}
d \left( \frac12 F \wedge F \right) = \tilde{q} \, \mathbb{Y}^{(5)}_{\tilde{\mathcal C}} = \tilde{q} \, d \Theta^{(4)}_{\tilde\Omega_2} \, .
\end{equation}
  The insertion of such an operator in the path integral corresponds to turning on a $B^{(2)}_{\mathcal{D}} \wedge B^{(2)}_{\mathcal{D}} = \tilde{q} \,  \Theta^{(4)}_{\tilde\Omega_2}$.\footnote{It satisfies 
$\oint_{S^4} B^{(2)}_{\mathcal{D}}\wedge B^{(2)}_{\mathcal{D}} = 2 \pi \tilde{q}$
where the four-sphere is embedded into the five-dimensional surface transverse to $\tilde{\mathcal C}$ surrounding the defect. }

Similarly, the insertion of the symmetry generator amounts to turning on the background gauge field $B^{(2)}_{CW} = \a_2 \mathbb{Y}^{(2)}_{\Sigma_4}$. Therefore, $\Big\langle U(\Sigma_4, \a_2) \, \mathcal{D}(\tilde{\mathcal C}) \Big\rangle = Z[ B^{(2)}_{\mathcal{D}}\wedge B^{(2)}_{\mathcal{D}}, 0 , \a_2 \mathbb{Y}^{(2)}_{\Sigma_4}]$.

As in the previous case, performing a topological $\lambda^{(1)}_{CW}$ gauge transformation \eqref{eq:transfs1} to get rid of the background gauge field $B^{(2)}_{CW}$, an anomalous contribution arises, due to the lack of conservation of the $J^{(4)}$ current in the presence of the defect. The `t Hooft anomaly correctly reconstructs the operator charge according to the following identity
\begin{equation}
    \label{GcwG3}
\Big\langle U(\Sigma_4, \a_2) \, \mathcal{D}(\tilde{\mathcal C}) \Big\rangle =
e^{i \a_2 \tilde{q} \int  \Theta^{(4)}_{\tilde\Omega_2}\wedge Y^{(2)}_{\Sigma_4}}
\Big\langle \mathcal{D}(\tilde{\mathcal C}) \Big\rangle \, .
\end{equation}
Once more, the charge is related to the linking number between $\tilde{\mathcal C}$ and $\Sigma_4$. 

This analysis can be generalized straightforwardly to $n$ dimensions, with $n \geq 3$. For $n >6$, new conserved currents of the form \eqref{CWA} must be considered. Objects charged under a generic $J^{(2k)}$ current are operators supported on $(n-2k-1)$-dimensional hypersurfaces. Their support can be realized as a $(n-2k-1)$-dimensional intersection of $k$ monopole surfaces $\Sigma_{n-3}$. In the path integral, they are then sourced by the product of $k$ factors, $B^{(2)}_{\mathcal{D}}\wedge \dots \wedge B^{(2)}_{\mathcal{D}}$.

\section{Super-Higher-Form Symmetries}\label{sec:supersymmetries}

In this section, we generalize the previous construction of higher-form symmetries and corresponding charged operators to the case of theories with supersymmetry. We will formulate the problem in a supergeometry framework by promoting
a generic supersymmetric field theory to be defined on a supermanifold. We stick to notations and conventions of \cite{Cremonini:2020mrk,Cremonini:2020zyn}, where more details on the construction of superforms in supermanifolds can be found.   

We consider a field theory defined in the supermanifold $\mathcal{SM}$, with dimension $(n|m)$,
where $n$ is the bosonic dimension and $m$ is the fermionic one. Locally, 
it is isomorphic to a superspace $\mathbb{R}^{(n|m)}$ parametrized by coordinates $z^A = (x^a, \theta^\alpha)$, where $a=0,\dots,n-1, \alpha=1,\dots,m$. 
Flat supervielbeins are given by $V^a = dx^a + \theta^\a \gamma^a_{\a\b} d\theta^\b$, $\psi^\a = d\theta^\a$, for a suitable choice of Clifford gamma matrices. 
The superdifferential $d$ is defined as
$d = V^a \partial_a + \psi^\alpha D_\alpha$, 
where $D_\alpha$ are the supercovariant derivatives satisfying $\{D_\alpha, D_\beta\} = -\gamma^a_{\alpha\beta} \partial_a$. From the identity $d V^a = \psi \gamma^a \psi$ it follows $d^2=0$.

The study of super higher-form symmetries in supermanifolds was initiated in \cite{Cremonini:2020zyn}. Here, we generalize that construction to include different types of super-higher-form currents and provide a large spectrum of explicit examples.  

In $\mathcal{SM}$, we define the $(n-p-1|q)$-form supercurrent $J^{(n-p-1|q)}$ to be a $(n-p-1|q)$-form, where the first index $(n-p-1)$ denotes the form degree and $q$ denotes the picture (number of Dirac delta functions $\delta(\psi)$\footnote{We recall that the delta's $\delta(\psi)$ carry no form degree, but they carry a new quantum number: 
the {\it Picture Number}, terminology directly imported form string-theory jargon. Note also that 
the derivatives $\partial/\partial \psi^\beta$ reduce the form degree since they act as the contraction operator
$\iota_{\alpha}$ with respect to an odd vector field $\partial/\partial \theta^\alpha$. }). We will  consider the two extreme cases $q=0$ and $q=m$, the former corresponding to a conserved {\it superform} current $J^{(n-p-1|0)} \equiv \star J^{(p+1|m)}$, the latter being a conserved  {\it integral form} current $J^{(n-p-1|m)} \equiv  \star J^{(p+1|0)}$ \footnote{Here $\star: \Omega^{(p|q)} \rightarrow  
 \Omega^{(n-p|m-q)}$ stands for the Hodge dual operator in supermanifolds \cite{Castellani:2015ata, Castellani:2016ezd}.}. More generally, one could consider $(n-p-1|q)$-form currents, for $q \neq 0,m$. This represents 
a very rich sector of new symmetries; nonetheless, defining these forms requires introducing a different geometric framework. Therefore, it goes beyond the scope of the present paper. We plan to come back to this interesting program in the future. 

The two sets of currents can be generally written as expansions in powers of the supervielbeins, precisely
\begin{equation}
\label{eq:superform}
J^{(n-p-1|0)}=  \sum_{k=0}^{n-p-1} J^{(0)}_{a_1 \dots a_k \, \a_{k+1} \dots \a_{n-p-1}} \, V^{a_1} \wedge \dots V^{a_k} \wedge  \psi^{\a_{k+1}}  \wedge \dots \psi^{\a_{n-p-1}} \, , 
\end{equation}
and
\begin{equation}
\label{eq:integralform}
J^{(n-p-1|m)}= \sum_{k=n-p-1}^n J^{(m) \quad \beta_1 \dots \beta_{k - n + p +1}}_{a_1 \dots a_k}  \, V^{a_1} \wedge \dots V^{a_k} \wedge \iota_{\b_1} \dots 
\iota_{\b_{k-n+p+1}}
\delta^{(m)} (\psi)  \, ,
\end{equation}
where $\iota_\b \equiv \frac{\partial}{\partial \psi^{\b}}$. The coefficients of the expansions are local superfields, functions of bosonic and fermionic coordinates. For example, the expansion 
of $J^{(1|2)}$ reads 
\begin{eqnarray}
    J^{(1|2)} = J_a V^a \delta^2(\psi) + J_{ab}^\alpha  V^a \wedge V^b \wedge \iota_\alpha\delta^2(\psi) + 
    J_{abc}^{\alpha\beta}  \, V^a \wedge V^b \wedge V^c \wedge \iota_\alpha \iota_\beta\delta^2(\psi) \, .
\end{eqnarray}

Ordinary vector superform currents in superspace correspond to $J^{(n-1|m)} \equiv \star J^{(1|0)}$, whose expansion is given in \eqref{eq:integralform} for $p=0$. For $p=1$, the closure of $J^{(n-2|m)}$ gives rise to the standard conservation of the U(1) supercurrent in superspace, $D_\a (J^\a + \gamma_a^{\a\b} D_\b J^a) = 0$ \cite{Cremonini:2020zyn}. This is the lowest representative of a chain of generalized conserved supercurrents obtained by increasing the form degree $p$ as in the ordinary bosonic case and/or changing the picture number. 

According to the general theory of integration on supermanifolds \cite{Witten:2012bg}, it is possible to give a precise meaning to the following integrals
\begin{eqnarray}
    \label{preB}
     Q(\Sigma_{(n-p-1|0)}) =\int_{\Sigma_{(n-p-1|0)}} J^{(n-p-1|0)}\,, ~~~~~~
     Q(\Sigma_{(n-p-1|m)}) = \int_{\Sigma_{(n-p-1|m)}} J^{(n-p-1|m)} \,,
\end{eqnarray}
where $\Sigma_{(n-p-1|0)}$, $\Sigma_{(n-p-1|m)}$ are non-contractible cycles in the supermanifold. They are parame\-tri\-zed as follows 
\begin{eqnarray}
    \label{preC}
    \Sigma_{(n-p-1|0)}&:& \quad  \mathbb{R}^{n-p-1} \rightarrow \SSM\,, ~~~~~~~~~~~~~     
    \Sigma_{(n-p-1|m)}: \quad \mathbb{R}^{(n-p-1|m)} \rightarrow \SSM\,, 
\\
    &~& ~\, \qquad \quad  t^i \rightarrow (x^a(t^i), \theta^\alpha(t^i)) \nonumber
    \,,  
    ~~~~~~~~~~~~~ \quad \qquad \quad (t^i, \eta^\beta) \rightarrow (x^a(t^i,\eta^\beta), \theta^\alpha(t^i,\eta^\beta) )  \, .  
\end{eqnarray}
The first cycle $\Sigma_{(n-p-1|0)}$ is a bosonic hypersurface embedded into the supermanifold, with the $t^i$-dependent fermionic coordinates describing the embedding. The second cycle $\Sigma_{(n-p-1|m)}$ is a super-hypersurface embedded into the supermanifold, which extends along all the fermionic directions.  

Both integrals in \eqref{preB} can be rewritten as integrals on the entire supermanifold using Poincar\'e duals. Precisely,
\begin{equation}
    \label{preD}
     Q(\S_{(n-p-1|0)}) =\int_{\SSM} J^{(n-p-1|0)}\wedge \mathbb{Y}^{(p+1|m)}_{\Sigma_{(n-p-1|0)}}\,, ~~~~~~~~
     Q(\S_{(n-p-1|m)}) =\int_{\SSM} J^{(n-p-1|m)} \wedge \mathbb{Y}^{(p+1|0)}_{\Sigma_{(n-p-1|m)}}\, , ~~~~~~~      
\end{equation}
where $\mathbb{Y}^{(p+1|m)}_{\Sigma_{(n-p-1|0)}}$ is an integral form  dual to the bosonic hypersurface $\Sigma_{(n-p-1|0)}$, whereas $\mathbb{Y}^{(p+1|0)}_{\Sigma_{(n-p-1|m)}}$ is a superform. 

Any change in the embedding of the $\Sigma$ cycles inside $\SSM$ is parametrized by a shift of the corresponding Poincar\' e dual form by an exact term, 
\begin{eqnarray}
    \delta \, \mathbb{Y}^{(p+1|m)}_{\Sigma_{(n-p-1|0)}} = d \Lambda^{(p|m)}\,, ~~~~~
    \delta \, \mathbb{Y}^{(p+1|0)}_{\Sigma_{(n-p-1|m)}} = d \Lambda^{(p|0)}\,. ~~~~~
\end{eqnarray}
It follows that on a supermanifold without boundary, if $\star J^{(p+1|m)}, \star J^{(p+1|0)}$ are closed forms (higher-superfom conserved currents), the integrals in \eqref{preB} define topological quantities. 
They correspond to the generators of a $(p|m)$-integral form and a $(p|0)$-superform symmetry, respectively, whose parameters $\lambda^{(p|q)}$ are
now an integral form if $q=m$ or a superform  if $q=0$.

Generalizing the construction of operators charged under bosonic higher-form symmetries, we expect to introduce two types of operators, the ones charged under $(p|0)$-superform symmetries \cite{Cremonini:2020zyn}
\begin{eqnarray}
\label{chA}
W(\Sigma_{(p|0)}) = \exp \left( i q \int_{\Sigma_{(p|0)}}A^{(p|0)} \right) = \exp \left( i q \int_{{\SSM}} A^{(p|0)} \wedge \mathbb{Y}^{(n-p|m)}_{\Sigma_{(p|0)}} \right)\, ,
\end{eqnarray}
and the ones charged under $(p|m)$-integral form symmetries
\begin{eqnarray}
\label{chA2}
\Tilde{W}(\Sigma_{(p|m)}) = \exp \left( i \tilde{q} \int_{\Sigma_{(p|m)}} A^{(p|m)} \right) = \exp \left( i \tilde{q} \int_{{\SSM}} A^{(p|m)} \wedge \mathbb{Y}^{(n-p|0)}_{\Sigma_{(p|m)}}\right) \, . 
\end{eqnarray}

The first set of operators was studied in \cite{Cremonini:2020zyn}. They carry a non-vanishing charge under the action of $Q(\Sigma_{[n-p-1|m]})$ in \eqref{preD}. Precisely, it was found that
\begin{eqnarray}
\label{eq:transfs}
    && e^{i\a \, Q(\S_{(n-p-1|m)})} \, W \, e^{-i\a \, Q(\S_{(n-p-1|m)})} = e^{i\a \, {\rm SLink}\left(\Sigma_{(p|0)},   \Sigma_{(n-p-1|m)}\right) } \, W  \, .
\end{eqnarray}
Here ${\rm SLink}$ is the super-linking number between the supersurfaces $\Sigma_{(p|0)}$ and $\Sigma_{(n-p-1|m)}$ \cite{Cremonini:2020zyn} (see also appendix \ref{app:linking}), 
\begin{eqnarray}
\label{chC}
{\rm SLink}\left(\Sigma_{(p|0)},  \Sigma_{(n-p-1|m)}\right)  
\! = \! \int_{{\SSM}} \! \! \! \! \Theta^{(n-p-1|m)}_{\Omega_{(p+1|0)}} \wedge  \mathbb{Y}^{(p+1|0)}_{\Sigma_{(n-p-1|m)}}
= (-1)^{n-p} \!\! \int_{{\SSM}}  \!\! \! \!
\mathbb{Y}^{(n-p|m)}_{\Sigma_{(p|0)}} \wedge \Theta^{(p|0)}_{\Omega_{(n-p|m)}} \, ,
\end{eqnarray} 
with 
$ \mathbb{Y}^{(n-p|m)}_{\Sigma_{(p|0)}} = d\Theta^{(n-p-1|m)}_{\Omega_{(p+1|0)}}$ ($\partial \Omega_{(p+1|0)} = \Sigma_{(p|0)}$) and 
$ \mathbb{Y}^{(p+1|0)}_{\Sigma_{(n-p-1|m)}} = d\Theta^{(p|0)}$ ($\partial \Omega_{(n-p|m)} = \Sigma_{(n-p-1|m)}$).
The two expressions in \eqref{chC} are equivalent up to integration by parts. 

Transformations \eqref{eq:transfs} correspond to the following shift of the connection superform
\begin{eqnarray}
A^{(p|0)} \rightarrow  A^{(p|0)} + \frac{\a}{q} \, \Theta^{(p|0)}_{\Omega_{(n-p|m)}} \, , 
\end{eqnarray}
which cannot be compensated by a supergauge transformation 
$\delta A^{(p|0)} = d C^{(p-1|0)}$. 

By analogy, we expect operators \eqref{chA2} to be charged under the action of $Q(\Sigma_{(n-p-1|0)})$ in \eqref{preD} with a charge proportional to the super-linking number ${\rm SLink}\left(\Sigma_{(p|m)},  \Sigma_{(n-p-1|0)}\right)$. 

In the next section, we will derive \eqref{eq:transfs} and its analog for $\tilde{W}$ using a path integral approach. We will make this construction concrete by studying several examples of super-Maxwell theories in various dimensions and with different amounts of supersymmetry. 

\section{Super-Maxwell theory}\label{sec:superMaxwell}

In a generic $(n|m)$-dimensional supermanifold $\SSM$, the super-Maxwell theory \cite{Brink:1976bc} is described in terms of an elementary $(1|0)$-superform gauge potential $A$, which can be expanded on the supervielbein basis as $A = A_a V^a + A_\a \psi^\a$, with $A_a, A_\a$ being local superfields. Its field strength 
$F = d A$ is the $(2|0)$-superform
\begin{eqnarray}
    \label{SIB}
    F = F_{ab} V^a \wedge V^b + F_{a\alpha} V^a \wedge \psi^\alpha + F_{\alpha\beta} \, \psi^\alpha \wedge \psi^\beta \, .
\end{eqnarray}
It is invariant under supergauge transformations, 
$\delta A = d \Lambda$, and satisfies the Bianchi identity $d F = 0$.  Redundant degrees of freedom in $F$ are removed by imposing suitable gauge invariant constraints (for example, the conventional constraint $F_{\alpha\beta} = 0$). 

The super-Maxwell action reads 
\begin{equation}
\label{eq:superaction}
    S = -\frac12\int_{\mathcal{SM}} F \wedge \star F \, , 
\end{equation}
and gives the equations of motion $d \star F = 0$. 
We recall that the $\star$ super-Hodge dual operator requires the definition of a supermetric in the supermanifold. A convenient way to define the spinorial components of such a metric is to couple the theory to a supergravity background \cite{Castellani:2023tip}. In general, this leads to higher-derivative field equations. The equations of motion for the physical degrees of freedom of super-Maxwell are eventually obtained by sending the background to zero. Since this procedure is dimension-dependent, we will properly discuss it case by case in the next sections. 

The super-Maxwell theory has two global super-higher-form symmetries generated by the Noether-like supercurrent $J^{(n-2|m)} = \star F$ and the topological supercurrent $J^{(2|0)} = F$. Depending on the specific dimension $n$, further topological symmetries can be considered, which are generated by the conserved superform currents 
\begin{eqnarray}
    \label{superCW}
    J^{(2k|0)} = \frac{1}{k!} \, \underbrace{F \wedge F \wedge \dots F}_k \, , \qquad \quad 2k < n \, .
\end{eqnarray}
We will call them super-Chern-Weil (super-CW) symmetries. 

However, this is not the end of the story. As we will see in a few examples, for suitable $(n|m)$ dimensions, other closed topological forms $\omega$ can be constructed in terms of the supervielbeins. These other forms can be combined with $J^{(2k|0)}$ to generate extra conserved currents of the form 
$\omega \wedge J^{(2k|0)}$. Since their construction involves objects that feature the geometry of the supermanifold, we will call them super-geometric-Chern-Weil (super-gCW) symmetries.

For many purposes, including the study of the action of symmetry generators on charged operators, it is convenient to couple the super-higher-form symmetries to background gauge fields. Therefore, the functional quantization of the theory will be described by the following path integral  
\begin{eqnarray}
     \label{pathint}
      Z[B^{(2|0)}, B^{(n-2k|m)}_{CW}, \dots] = \int [\mathcal{D} A] \, e^{ i \, ( S +S_{min} + S_{g.f.})} \, ,
 \end{eqnarray}
where $S$ is defined in \eqref{eq:superaction}, $S_{g.f.}$ is a suitable gauge-fixing action and $S_{min}$ gives the minimal coupling with the background gauge superforms,
\begin{eqnarray}
    \label{eq:Smin}
    S_{min} = \int_{\SSM} \left( J^{(n-2|m)}\wedge B^{(2|0)} + \sum_k J^{(2k|0)} \wedge B^{(n-2k|m)}_{CW}  + \dots  
    \right) \, .
\end{eqnarray}
Dots stand for extra couplings with more general conserved currents of the form $\omega \wedge J^{(2k|0)} $.

\subsection{ $D=(3|2)$}

As a warming up, we consider the simple case of super-Maxwell theory in three dimensions, with $N=1$ supersymmetry. 

We parametrize the $D=(3|2)$ superspace with coordinates $z^A = (x^a, \theta^\a)$, where in Lorentz signature $a=0,1,2$, whereas $\a = 1,2$ labels the spinorial representation of SL(2,$\mathbb R$). We introduce superspace covariant derivatives $D_\a$, satisfying the supersymmetry algebra $\{ D_\a , D_\b \} = -\partial_{\a\b}$
\footnote{After assigning a set $\{ \sigma^a \}$ of $2 \times 2$ Clifford matrices, for $v$ vectors we use the double index notation 
$v^{\a\b} = (\sigma^a)^{\a\b} v_a$.}, and the superdifferential $d = V^{\a\b} \partial_{\a\b} + \psi^\a D_\a$, written in terms of superdreibeins \cite{Castellani:2015paa}
\begin{equation}
    V^{\a\b} = dx^{\a\b} + \theta^\a d 
    \theta^\b \, , \qquad \psi^\a = d\theta^\a \, .
\end{equation}
They satisfy $dV^{\a\b} = \psi^\a \wedge \psi^\b$, $d \psi^\a = 0$.

The $(2|0)$-superform $F$ can be expanded on the superdreibeins basis, as  
\begin{eqnarray}
\label{S3A}
 F = V^a \wedge V^b \, F_{ab} + \psi^{\alpha} \wedge V_{\alpha \beta} \, W^\beta 
+ \psi^\a \wedge \psi^\b F_{\a\b} \, ,\end{eqnarray}
where $F_{ab}, F_{\a\b}$ and $W^\beta$ are the field strength and the gluino superfields, respectively. Requiring $dF=0$ and the conventional constraint $F_{\a\b}=0$, the following Bianchi identities follow
\begin{eqnarray}
\label{S3B}
D_\alpha W^\alpha =0\,,~~~~~ F_{ab} = D^\a (\sigma_{ab})_{\a\b}  W^\b \, .
\end{eqnarray}

The super-Hodge dual in the $(3|2)$-supermanifold has been constructed in \cite{Castellani:2015paa,
Castellani:2015ata}. According to that construction, we have (we list only the identities relevant to our discussion)
\begin{eqnarray}
\label{S3C}
&&\star ( V^a \wedge V^b) = \epsilon^{ab}_{\; \; \; \; c} \, V^c \wedge \delta^{(2)}(\psi) \,, \nonumber \\
&& \star (\psi^\alpha \wedge V^a) = \epsilon^a_{\; \; bc} V^b \wedge V^c \, \wedge \iota^\alpha \delta^{(2)}(\psi) \, ,
\nonumber \\
&& \star (\psi^\alpha \wedge \psi^\b) = \epsilon_{abc} V^a \wedge V^b \wedge V^c \wedge \iota^\alpha \iota^\b \delta^{(2)}(\psi) \, ,
\end{eqnarray}
where $\iota_\alpha \equiv \partial/\partial \psi^\a$.  From these identities, the $(1|2)$-integral form dual to the superfield strength follows, which reads
\begin{eqnarray}
\label{S3D}
\star F &=& F^{ab} \epsilon_{abc} V^c \wedge \delta^{(2)}(\psi) + 
W^\alpha \sigma^a_{\alpha \beta}\epsilon_{abc} V^b \wedge V^c \wedge \iota^\beta \delta^{(2)}(\psi) \, .
\end{eqnarray}
The equations of motion $d \star F =0$ derived from the action \eqref{eq:superaction}, together with Bianchi identities \eqref{S3B} are equivalent to the Maxwell and Dirac equations in superspace 
\begin{eqnarray}
\label{S3E}
\partial_{(\alpha}^{~~\beta} f_{\beta|\gamma)} =0\,,~~~~ \partial^{\alpha\beta}W_\beta =0 \, , 
\end{eqnarray}
and the Bianchi identities 
\begin{eqnarray}
\nonumber 
\partial^{\alpha\beta} f_{\alpha\beta} =0 \, ,
\end{eqnarray}
where $f_{\alpha\beta} \equiv (\sigma^{ab})_{\alpha\beta} F_{ab}$.

\paragraph{Supersymmetry currents.} Ordinary supertranslations in superspace correspond to $0$-form symmetries implemented by closed $(2|2)$-form currents with external spinorial and vectorial indices, respectively, given by
\begin{eqnarray}
\label{susycurr}
J^{(2|2)}_\alpha &=& \psi_\a \wedge J^{(1|2)}\, , \qquad {\rm with}  \quad   J^{(1|2)} = (V \wedge V)^{\b\g} f_{\d\b}W_\g \wedge \iota^\d \delta^{(2)}(\psi)   \, ,  \\
J^{(2|2)}_{(\alpha\b)} &=&
dV_{\a\b} \wedge  
J^{(0|2)} \, , \quad {\rm with} \quad
J^{(0|2)} =
(V \wedge V)^{\g\d}\left(  f_{\epsilon\g} f_{\eta\delta} 
+ W_\g \partial_{  \epsilon\eta} W_\d
\right) \wedge  \iota^\epsilon \iota^\eta \delta^{(2)}(\psi)  \, , \nonumber
\end{eqnarray}
where $(V \wedge V)^{\b\g} \equiv V^{\b\delta} \wedge V_\delta^{\, \, \g}$. They belong to the class of closed currents of the form $\omega \wedge J$, where $dJ=0$ and $\omega$ is a closed form constructed in terms of the superdreibeins. 
Integrating by parts the $\iota$-derivatives in \eqref{susycurr}, we obtain their explicit expressions
\begin{equation}
J^{(2|2)}_\alpha = (V \wedge V)^{\b\g} f_{\a\b}W_\g \,\delta^{(2)}(\psi)     \, , \qquad 
J^{(2|2)}_{(\alpha\b)} = (V \wedge V)^{\g\d}\left(  f_{\a\g} f_{\b\delta} 
+ W_\g \partial_{  \alpha \b} W_\d
\right) \delta^{(2)}(\psi)  \, .
\end{equation}
Their closure
is ensured by the Dirac equation of motion, the Bianchi identities $D_\alpha f_{\beta\gamma} = 
\partial_{\alpha (\beta} W_{\gamma)}$ and 
$f_{\alpha \beta} = D_\alpha W_\beta$ 
and the equations of motion in \eqref{S3E}. Moreover, $\partial^{\alpha\b} J^{(2|2)}_{(\alpha\b)} =0$ on-shell. 

It is instructive to discuss how many conservation laws follow from the closure of these currents. To this end, we consider a generic $(1|2)$-integral form expanded as in \eqref{eq:integralform}, 
\begin{equation}
J^{(1|2)} = J_{\a\b} V^{\a\b} \delta^2(\psi) + 
J^{(\a\b) \g} (V\wedge V)_{\a\b} \iota_\g \delta^2(\psi) + \hat J^{\a\b} (V\wedge V\wedge V) \iota_\a \iota_\b \delta^2(\psi)\, .
\end{equation}
Imposing the closure condition $dJ^{(1|2)}=0$ then leads to a set of two equations
\begin{equation}
\partial_{(\a |\g} J^\g_{~\b)} + D_\gamma J_{(\a\b)}^{~~\g} + \hat J_{(\a\b)} =0 \, , \qquad 
 \partial_{(\a\b)} J^{(\a\b)\g} + D_\b \hat J^{\g\b}=0 \, . 
\end{equation}
The first equation allows to express $\hat J_{(\a\b)}$ in terms of the other components. The second equation, instead, reproduces the ordinary conservation law in superspace, 
\begin{equation}
D_\b (D_\a J^{(\a\b) \g} + \hat J^{\g\b}) = 0 \, .  \end{equation}
Similarly, for a generic $J^{(0|2)}$ integral form current we obtain two independent conservartion laws. 

In the present framework, the generators associated to the supersymmetry and translation currents in \eqref{susycurr} are given by
\begin{eqnarray}
    \label{newSUSA}
    Q_\a =  \int_{\Sigma_{(2|2)}} J^{(2|2)}_\alpha\,, ~~~~~~\quad 
    Q_{(\a\b)} = 
    \int_{\Sigma_{(2|2)}} J^{(2|2)}_{(\alpha\b)}\,. ~~~~~~
\end{eqnarray}
where $\Sigma_{(2|2)}$ is a supersurface embedded in $\SSM$. 

An extended analysis of supercurrents in the framework of supergeometry will be presented elsewhere \cite{grassi}. 

\paragraph{Electric and magnetic currents.} Beyond ordinary supertranslation invariance in superspace, the 3D super-Maxwell theory enjoys two extra symmetries, a Noether-like $(1|0)$-superform symmetry generated by $\star F$ and a topological $(0|2)$- integral form symmetry generated by $F$. The corresponding generators are the following topological operators 
\begin{equation}
\label{S3F}
Q_e(\Sigma_{(1|2)}) =  \int_{\SSM}  \star F
\wedge \mathbb{Y}^{(2|0)}_{\Sigma_{(1|2)}}\, , \qquad \qquad
Q_m(\Sigma_{(2|0)}) =  \int_{\SSM}  F \wedge \mathbb{Y}^{(1|2)}_{\Sigma_{(2|0)}} \, , 
\end{equation}
where $\Sigma_{(1|2)}$ is a superline and  $\Sigma_{(2|0)}$ an ordinary surface in $\SSM$. 
The integrals have been expressed in terms of their Poincar\'e duals. 

In order to determine $\mathbb{Y}^{(2|0)}_{\Sigma_{(1|2)}}$, 
we need to establish how the surface $\Sigma_{(1|2)}$ is immersed into 
the supermanifold $\SSM$. 
This is given by two defining equations $\Phi_1=0$ and 
$\Phi_2=0$ (with ${\rm Jac}(\Phi_1, \Phi_2) \neq 0$), where in general $\Phi_I$ are functions of the bosonic and fermionic coordinates. 
Generalizing the definition \eqref{eq:PCO} for Poincar\'e duals in ordinary manifolds, we write
\begin{eqnarray}
    \label{new_PCOA}    \mathbb{Y}^{(2|0)}_{\Sigma_{(1|2)}} = \delta(\Phi_1) d\Phi_1 \wedge 
    \delta(\Phi_2) d\Phi_2 \, .
\end{eqnarray}

Analogously, the Poincar\'e dual $\mathbb{Y}^{(1|2)}_{\Sigma_{(2|0)}}$ is defined by the 
immersion $\Sigma_{(2|0)} \hookrightarrow \SSM$. However, in this case, we have to introduce
 one bosonic equation $\Phi=0$ and two independent fermionic ones
$\Theta_\a =0$, $
\a=1,2$, needed to localize the fermionic coordinates. The Poincar\'e dual is then given by the following $(1|2)$ integral form
\begin{eqnarray}
     \label{new_PCOB}   \mathbb{Y}^{(1|2)}_{\Sigma_{(2|0)}} = 
    \delta(\Phi) d\Phi \bigwedge_{\a=1}^2 \Theta_\a \delta(d \Theta_\a) \, .
\end{eqnarray}
We have used $\delta( \Theta_\a) =  \Theta_\a$, valid for fermionic functions. 

It is easy to check that $ \mathbb{Y}^{(2|0)}_{\Sigma_{(1|2)}} $ and $ \mathbb{Y}^{(1|2)}_{\Sigma_{(2|0)}} $ belong to the cohomology classes $H^{(2|0)}(d)$ and $H^{(1|2)}(d)$, respectively. 
Moreover, any variation of the embedding of $\Sigma_{(1|2)}$ and $\Sigma_{(2|0)}$ into the supermanifold amounts to adding a d-exact term  to $\mathbb{Y}^{(2|0)}_{\Sigma_{(1|2)}}$ and $ \mathbb{Y}^{(1|2)}_{\Sigma_{(2|0)}} $. Due to the closure of the currents, in the absence of non-trivial boundaries, this does not affect expressions \eqref{S3F} for the charges, which are then topological operators.   

In three dimensions, there are no extra supercurrents of the form \eqref{superCW}. Therefore, coupling the $\star F$ and $F$  currents to a background gauge superform and an integral form, respectively, the total action $S+ S_{min}$ appearing in the path integral \eqref{pathint} acquires a very simple form
\begin{equation}
    \label{eq:Smin3d}
    S+ S_{min} = \int_{\SSM} \left( - \frac12 (F - B^{(2|0)}) \wedge \star (F - B^{(2|0)}) +  F \wedge B_{CW}^{(1|2)} \right) \, .
\end{equation}

This action is invariant under the following gauge transformations
\begin{eqnarray}
\label{eq:transfs1A}
    F \rightarrow F + d\Lambda^{(1|0)} 
    \, , \qquad  B^{(2|0)} \rightarrow B^{(2|0)} + d \Lambda^{(1|0)} \, , \qquad 
    B^{(1|2)}_{CW} \rightarrow B^{(1|2)}_{CW} + d \Lambda^{(0|2)} \, ,
\end{eqnarray}
up to an anomaly term given by 
\begin{equation}
\label{SGcwE2}
    S' + S_{min}' = S + S_{min} 
    + \int_{\SSM}  d\Lambda^{(1|0)} \wedge B^{(1|2)}_{CW} \, .
\end{equation}
Gauge transformations are generated by gauge parameters 
$\Lambda^{(1|0)}$ and  $\Lambda^{(0|2)}$ that are superform and integral forms, respectively. 

The operator charged under $Q_e$  is the superspace generalization of the Wilson loop \cite{Cremonini:2020mrk,Cremonini:2020zyn}. This can be equivalently written in terms of the superconnection $A$ or its superfield strength $F =dA$ as
\begin{eqnarray}
\label{STSL0}
W(\mathcal{C}) = e^{i q_e \Gamma } \,, ~~~~~
\Gamma = \int_{\SSM} A \wedge \mathbb{Y}^{(2|2)}_\mathcal{C} = \int_{\SSM} F \wedge \Theta^{(1|2)}_{\Omega_{(2|0)}} \, ,
\end{eqnarray}
where $\Omega_{(2|0)}$ is a $(2|0)$ surface whose boundary is the $(1|0)$ closed line $\mathcal{C}$. Consequently,  the corresponding Poincar\'e duals satisfy $\mathbb{Y}^{(2|2)}_{\mathcal C} = d \Theta^{(1|2)}_{\Omega_{(2|0)}}$. 

 The insertion of $W(\mathcal{C})$ inside the path integral corresponds to turning on the background $B^{(1|2)}_{CW} = q_e \Theta^{(1|2)}_{\Omega_{(2|0)}}$ in \eqref{eq:Smin3d}. Analogously, the insertion of the generator $U(\Sigma_{(1|2)}, \a_e) 
 = e^{i\a_e Q_e(\Sigma_{(1|2)})}$ amounts to turning on the background $B^{(2|0)} = \a_e \mathbb{Y}^{(2|0)}_{\Sigma_{(1|2)}}$. Therefore, we can write
 \begin{equation}
  \Big\langle U(\Sigma_{(1|2)}, \alpha_e) \, W(\mathcal{C}) \Big\rangle =
    Z[\alpha_e \mathbb{Y}^{(2|0)}_{\Sigma_{(1|2)}}, \, q_e \Theta^{(1|2)}_{\Omega_{(2|0)}}] \, .
 \end{equation}
 
In order to compute the  
charge of $W$ under the action of $Q_e$ we perform a gauge transformation with $\Lambda^{(1|0)} = - \a_e \mathbb{Y}^{(2|0)}_{\Sigma_{(1|2)}}$ to remove $B^{(2|0)}$. We are then left with $Z[0, q_e \Theta^{(1|2)}_{\Omega_{(2|0)}}] = \langle W(\mathcal{C}) \rangle$, times a phase due to the anomaly in \eqref{SGcwE2}. 
Inserting there our explicit choice for $\Lambda^{(1|0)}$ and $B^{(1|2)}_{CW}$ we eventually find 
\begin{equation}
  \Big\langle U(\Sigma_{(1|2)}, \alpha_e) \, W(\mathcal{C}) \Big\rangle =
    e^{i \a_e q_e \, {\rm SLink}(\Sigma_{(1|2)}, \, \mathcal{C}) } \Big\langle W({\mathcal C}) \Big\rangle \, ,
 \end{equation}
 where 
\begin{eqnarray}
\label{S3G2}
{\rm SLink}(\Sigma_{(1|2)}, \, \mathcal{C})  = \int_{\SSM} \mathbb{Y}^{(2|0)}_{\Sigma_{(1|2)}} \wedge \Theta^{(1|2)}_{\Omega_{(2|0)}} = 
 \int_{\SSM} \Theta^{(1|0)}_{\Omega_{(2|2)}}\wedge \mathbb{Y}^{(2|2)}_{\mathcal{C}} 
\end{eqnarray}
is the super-linking number \eqref{chC}. 
In the last equality we have defined $\mathbb{Y}^{(2|0)}_{\Sigma_{(1|2)}}  = d \Theta^{(1|0)}_{\Omega_{(2|2)}}$. 

We now move to construct operators charged under $Q_m$ defined in \eqref{S3F}. In analogy with the bosonic case, such operators are supermonopoles, that is, defects in the supermanifold that spoil the conservation law $d F=0$. 
As in the bosonic case, we split $F = F_r + F_s$, with 
\begin{equation}
  dF_r = 0 \; (\Rightarrow F = dA_r ) \, , \quad {\rm and } \quad dF_s = q_m \mathbb{Y}^{(3|0)}_{P_{(0|2)}} \, . 
\end{equation}
 Here, $\mathbb{Y}^{(3|0)}_{P_{(0|2)}}$ represents the Poincar\'e dual of a point-like defect carrying charge $q_m$. We dub $P_{(0|2)}$ a 't Hooft superpoint, as it is localized in the bosonic part of $\SSM$ but it extends along the superspace components.

Considering a superline $\Sigma_{(1|2)}$ whose boundary is the superpoint, we can write $\mathbb{Y}^{(3|0)}_{P_{(0|2)}} = d \Theta^{(2|0)}_{\Sigma_{(1|2)}}$. Therefore, the insertion of the superpoint in the path integral, which amounts to implementing the split of $F$ while integrating only on $A_r$, corresponds to turning on a background $B^{(2|0)}_{\mathcal{D}} = q_m\Theta^{(2|0)}_{\Sigma_{(1|2)}} $. 

To study the action on the defect of the symmetry generator $U(\Sigma_{(2|0)}, \a_m) = e^{i \a_m Q_m(\Sigma_{(2|0)})}$ with $Q_m(\Sigma_{(2|0)})$ given in \eqref{S3F}, we evaluate the two-point function
\begin{equation}
  \Big\langle   U(\Sigma_{(2|0)}, \a_m) \, \mathcal{D}(P_{(0|2)}) \Big\rangle = 
  Z[B^{(2|0)}_{\mathcal{D}}, \, B^{(1|2)}_{CW} = \a_m \mathbb{Y}^{(1|2)}_{\Sigma_{(2|0)}}] \, .
\end{equation} 
We perform a gauge transformation \eqref{eq:transfs1} with $d \Lambda^{(0|2)}_{CW} = - \a_m \mathbb{Y}^{(1|2)}_{\Sigma_{(2|0)}}$ to get rid of $B^{(1|2)}_{CW}$. In the presence of the 't Hooft superpoint, this transformation generates an anomaly, which eventually gives rise to the defect charge. Precisely, we obtain
\begin{equation}
  \Big\langle   U(\Sigma_{(2|0)}, \a_1) \, \mathcal{D}(P_{(0|2)}) \Big\rangle = 
  e^{i \a_1 q \, {\rm SLink}(P_{(0|2)}, \Sigma_{(2|0)})} 
  \Big\langle \mathcal{D}(P_{(0|2)}) \Big\rangle \, ,
\end{equation} 
where, according to the general definition in \eqref{chC}, the super-linking number is given by
\begin{equation}
   {\rm SLink}(P^{(0|2)}, \Sigma_{(2|0)}) = \int_{\SSM}  \Theta^{(2|0)}_{\Sigma_{(1|2)}} \wedge \mathbb{Y}^{(1|2)}_{\Sigma_{(2|0)}} \, .
\end{equation}

\subsection{$D= (4|4)$} 
\label{sec:4D}

A more relevant example is the 
$U(1)$ pure gauge theory in flat $N=1$ superspace. 

Assigning coordinates $z^A = (x^a, \theta^\a, \bar\theta^{\dot\a})$, $a = 0,1,2,3$, $\a, \dot\a = 1,2$, and equipping the superspace with supersymmetry covariant derivatives satisfying $\{ D_\a , \bar{D}_{\dot\a} \} = - \partial_{\a \dot\a}$ \footnote{For tensorial quantities we will use indifferently the vector or the double-index notation, with $v^a = (\sigma^a)^{\a\dot\a} v_{\a \dot\a}$.}, we define the super differential $d = V^{\a\dot\a} \partial_{\a \dot\a} + \psi^\a\ D_\a + \bar\psi_{\dot\a} \bar{D}^{\dot\a}$ \cite{Castellani:2023tip}, with the supervielbeins explicitly realized as   
\begin{equation}
    V^{\a \dot\a} = d x^{\a \dot\a} + \frac12 (\theta^\a d\bar\theta^{\dot\a} + d\theta^\a \bar\theta^{\dot\a}) \, \qquad \psi^\a = d\theta^\a \, , \qquad \bar\psi^{\dot\a} = d\bar\theta^{\dot\a} \, ,
\end{equation}
They satisfy $dV^{\a\dot\a} = \psi^\a \wedge \bar{\psi}^{\dot\a}$, $d\psi^\a = d\bar{\psi}^{\dot\a} = 0$.

Super-Maxwell theory in four dimensions 
has two bosonic and two fermionic on-shell degrees of freedom, which fit into the $(2|0)$-superform
\begin{equation}
\label{STSA}
 F = V^a \wedge V^b \, F_{ab} + \bar\psi_{\dot \alpha} \wedge V^{\alpha \dot\alpha} \, W_\alpha + 
   \bar W_{\dot \alpha} \, V^{\alpha \dot\alpha} \wedge  \psi_{\alpha}  +  \psi^\a \wedge \psi^\b F_{\a \b} + \bar\psi^{\dot\a} \wedge \bar\psi^{\dot\b} F_{\dot\a \dot\b} + \psi^\a \wedge \bar\psi^{\dot\b} F_{\a \dot\b}\, ,
\end{equation}
with the coefficients functions being field strength superfields. Conventional constraints correspond to set $F_{\a\b}=F_{\dot\a \dot\b} = F_{\a \dot\b}=0$. Moreover, we write $F_{ab} \equiv (\sigma_a)^{\a\dot\a} (\sigma_b)^{\b\dot\b} ( \epsilon_{\dot\a \dot\b} f_{\a\b} + \epsilon_{\a\b} \bar{f}_{\dot\a \dot\b})$.

$F$ is subject to the Bianchi identity $d F=0$, which explicitly gives 
\begin{eqnarray}
\label{STSB}
&&D^\alpha W_\alpha + \bar D_{\dot\alpha} \bar W^{\dot\alpha}=0\,,~~~~~
D_\alpha \bar W_{\dot\alpha}=0\,, ~~~~~~
\bar D_{\dot\alpha} W_\alpha = 0 \nonumber \\
&& 
f_{\alpha \beta} = D_{(\alpha} W_{\beta)}\,, ~~~~
\bar{f}_{\dot \alpha \dot \beta} = \bar D_{(\dot \alpha} \bar W_{\dot \beta)}\,. ~~~~
\end{eqnarray}

The Hodge dual of the superform $F$ can be evaluated using the general construction of the Hodge operator in supermanifolds given in \cite{Castellani:2015ata}. In particular, one obtains 
\begin{eqnarray}
\label{STSC}
\star (V^a \wedge V^b) &=& \epsilon_{abcd} V^b \wedge V^c \wedge \delta^{(4)}(\psi) \,, ~~~~~\nonumber \\
\star (\psi^\alpha \wedge V^b) &=& \epsilon_{abcd} (V^b \wedge V^c \wedge V^d)  \wedge \iota_\alpha \delta^{(4)}(\psi) \,, ~~~~~\nonumber \\
\star (\bar\psi^{\dot\alpha} \wedge V^b) &=& \epsilon_{abcd} (V^b \wedge V^c \wedge V^d) \wedge \bar\iota_{\dot\alpha}  \delta^{(4)}(\psi) \,, ~~~~~
\end{eqnarray}
where $\iota_\alpha \equiv \partial/\partial \psi^\alpha$ and $\bar\iota_{\dot\alpha} \equiv \partial/\partial \bar\psi^{\dot\alpha}$. 
Therefore, we find (removing the wedge symbol)
\begin{equation}
\label{STSD}
\star F = F^{ab} \epsilon_{abcd} V^b V^c \delta^4(\psi) + 
W^\alpha\gamma^a_{\alpha \dot \alpha}\epsilon_{abcd} (V^b V^c V^d) \bar\iota^{\dot\a} \delta^{(4)}(\psi)  
+
\bar W^{\dot\alpha} 
\gamma^a_{\alpha \dot \alpha} \epsilon_{abcd} (V^b V^c V^d) \iota_{\a} \delta^{(4)}(\psi) \, .
\end{equation}
This is a $(2|4)$ integral form. Taking into account the equations of motion from \eqref{eq:superaction}, $d \star F =0$, together with the Bianchi identities \eqref{STSB}, the coefficient superfields in the $\star F$ expansion satisfy
\begin{eqnarray}
\label{STSE}
\partial^a F_{ab}=0\,,~~~~ \partial^{\alpha\dot\alpha }W_\alpha =0 \,, ~~~~~
 \partial^{\alpha\dot\alpha }\bar W_{\dot\alpha} =0 \, .
\end{eqnarray}

\paragraph{Supersymmetry currents.} Conserved currents associated to supertranslation invariance in superspace and R-symmetry can be conveniently written in the present formalism. They correspond to the following closed $(3|4)$-integral form currents 
\begin{eqnarray}
    \label{cocEBC}
 J^{(3|4)} &=&  W_\a \bar W_{\dot \a} 
 (V\wedge V\wedge V)^{\alpha \dot\alpha} 
 \delta^{(4)}(\psi)  \\
J^{(3|4)}_\a &=& \psi_\a \wedge J^{(2|4)} \, , \qquad {\rm with} \qquad J^{(2|4)} = f_{\g\b} \bar W_{\dot\b} (V\wedge V\wedge V)^{\b \dot\b} \wedge \iota^\g \delta^{(4)}(\psi) \nonumber \\
\bar J^{(3|4)}_{\dot\a} &=& \bar{\psi}_{\dot\a} \wedge \bar{J}^{(2|4)} \, , \qquad {\rm with} \qquad \bar{J}^{(2|4)} = \bar{f}_{\dot\g\dot\b} W_{\b} (V\wedge V\wedge V)^{\b \dot\b} \wedge \iota^{\dot\g} \delta^{(4)}(\psi) \nonumber \\
 J^{(3|4)}_{\a\dot\a} &=& 
    dV_{\a\dot\a} \wedge J^{(1|4)} \, , \quad {\rm with} \qquad J^{(1|4)} = T_{ \b\dot\b\g\dot\g} 
    (V\wedge V\wedge V)^{\b \dot\b}  \iota^\g \iota^{\dot\g} \delta^{(4)}(\psi) \, ,
    \nonumber
\end{eqnarray}
where $T_{ \b\dot\b\g\dot\g} $ is the total energy-momentum tensor
\begin{equation}
T_{ \b\dot\b\g\dot\g} 
 = 2f_{\b\g} \bar{f}_{\dot\b\dot\g} - 
 \left(
    \bar W_{\dot \beta} \partial_{\g\dot\g} W_\b + \bar W_{\dot \g} \partial_{\b\dot\b} W_\g - 
    \partial_{\b\dot\b} \bar W_{\dot \g} W_\g - 
    \partial_{\g\dot \g} \bar W_{\dot \b} W_\b
    \right) \, ,
 \end{equation}
satisfying $\partial^{\b\dot\b} T_{\b\dot\b\g\dot\g} = 0$. The bosonic and fermionic components of $T$ are separately conserved, thanks to the equations of motion and Bianchi identities, $\partial^{\a\dot\a} W_\a = \partial^{\a\dot\a} \bar W_{\dot\a}=0, \partial^{\a\dot\a} f_{\alpha\beta} =0$ and 
$\partial^{\a\dot\a} \bar{f}_{\dot\alpha\dot\beta} =0$.

The currents in \eqref{cocEBC} are of the form $\omega \wedge J$, where $\omega$ are closed forms written in terms of the supervielbeins. They are responsible for assigning tensorial or spinorial nature to the supersymmetry and R-symmetry currents. 

The corresponding $0$-form symmetries are generated by the following topological charges
\begin{eqnarray}
\label{newSUB} 
&&Q_{\rm R} =  \int_{\Sigma_{(3|4)}}   J^{(3|4)}\,, ~~~~
Q_{\a\dot\a} = \int_{\Sigma_{(3|4)}}   J^{(3|4)}_{\alpha\dot \alpha}\,, ~~~~\\
&&Q_\a = \int_{\Sigma_{(3|4)}}   J^{(3|4)}_\alpha\,, ~~~~ \bar{Q}_{\dot\a} = 
\int_{\Sigma_{(3|4)}}   \bar{J}^{(3|4)}_{\dot\alpha}\,, ~~~~
\end{eqnarray}
for a given $\Sigma_{(3|4)}$ supersurface. 
As an illustrative example, we compute the  R-symmetry charge $Q_{\rm R}$ explicitly,
\begin{eqnarray}
    \label{CA_A}
    Q_{\rm R} = \int_{\Sigma_{(3|4)}} J^{(3|4)} = 
    \int_{\SSM}  J^{(3|4)}\wedge \mathbb{Y}^{(1|0)}_{\Sigma_{(3|4)}} \, ,
\end{eqnarray}
when $\Sigma_{(3|4)}$ is the unitary supersphere  $\mathbb{S}^{(3|4)} = OSp(4|4)/OSp(3|4)$, defined by the algebraic curve 
\begin{eqnarray}
    \label{CA_BA}
    \Phi \equiv x_{\alpha\dot\alpha}     x^{\alpha\dot\alpha} + \mu ( \theta^\alpha \theta_\alpha +  \bar\theta_{\dot\alpha} 
    \bar\theta^{\dot\alpha}) - 1 =0 \, .
\end{eqnarray}
Using the corresponding Poincar\'e dual form 
$\mathbb{Y}^{(1|0)}_{\Sigma_{(3|4)}} = \delta(\Phi) d\Phi$, we obtain 
\begin{eqnarray}
    \label{CA_BB}
    Q(\Sigma_{(3|4)})  &=& 
    \int_{\SSM}  x^{\alpha \dot \alpha} W_\alpha \bar W_{\dot \alpha}
    \, V^4 \wedge \delta^{(4)}(\psi) \, \delta(x_{\alpha\dot\alpha}     x^{\alpha\dot\alpha} + \mu ( \theta^\alpha \theta_\alpha +  \bar\theta_{\dot\alpha} 
    \bar\theta^{\dot\alpha}) - 1 )
    \nonumber \\
    &=& 
    \int  [d^4 x \, d^4 \theta] \; x^{\alpha \dot \alpha} W_\alpha \bar W_{\dot \alpha}
   \, \delta(x_{\alpha\dot\alpha}     x^{\alpha\dot\alpha} + \mu ( \theta^\alpha \theta_\alpha +  \bar\theta_{\dot\alpha} 
    \bar\theta^{\dot\alpha} ) - 1) \, ,
\end{eqnarray}
 where in the first line we have defined $V^4 = {\rm tr}(V \wedge V \wedge V \wedge V )$, and in the second line we have used $V^4 \wedge \delta^{(4)}(\psi) = [d^4x d^4\theta]$. The expression for $Q$ has been reduced to an ordinary superspace integral. It can be further simplified by performing the Berezin integral first. This requires expanding the Dirac delta function in powers of $\theta, \bar\theta$, thus obtaining a final expression that contains extra spacetime derivatives. Eventually, we find
 \begin{eqnarray}
    \label{CA_B}
    Q(\Sigma_{(3|4)})&=&  
 \int  d^4 x \, d^4 \theta \; x^{\alpha \dot \alpha} W_\alpha \bar W_{\dot \alpha}
   \, \Big( 
   \delta(x^2 - 1) + 
   \mu \left( 
    \theta^\alpha \theta_\alpha + \bar\theta_{\dot\alpha} 
    \bar\theta^{\dot\alpha}
   \right)  \delta'(x^2 - 1) 
   + 
   \left( 
 \mu^2  \theta^2  \bar\theta^2
   \right)  \delta''(x^2 - 1)  
    \Big) \nonumber \\
    &=&
\int  d^4 x \; x^{\alpha \dot \alpha} \left. D^2 W_\alpha\right| \left. 
\bar D^2\bar W_{\dot \alpha} \right|
   \, \delta(x^2 - 1) + 
2\mu \, x^{\alpha \dot \alpha} \Big( \left. W_\alpha\right| \left. 
\bar D^2\bar W_{\dot \alpha} \right| + 
 \left. D^2 W_\alpha\right| \left. 
\bar W_{\dot \alpha} \right| \Big)  \, \delta'(x^2 - 1) \nonumber 
\\
&+& 
\int  d^4 x \;
x^{\alpha \dot \alpha} (4 \mu^2 \left. W_\alpha\right| \left. \bar W_{\dot \alpha} \right|)  \, \delta''(x^2 - 1) \, ,
    \end{eqnarray}
where $\left. D^2 W_\alpha\right|\,, \dots, \left. \bar W_{\dot \alpha} \right|$ are components of the superfield $W_\a$ and $\bar W_{\dot \a}$. The last integral to be performed is over the Dirac delta 
$\delta(x^2 - 1)$ and its derivatives. 

\paragraph{Electric and magnetic currents.} We now turn to the two conserved superform currents $J^{(2|4)} = \star F $ and $J^{(2|0)} = F$. They generate a $(1|0)$-superform symmetry of Noether type and a $(1|4)$- integral form topological symmetry, respectively, whose generators can be written as 
\begin{equation}
\label{eq:4dcharges}
 Q_e(\Sigma_{(2|4)}) = \int_{\SSM} \star F \wedge \mathbb{Y}^{(2|0)}_{\Sigma_{(2|4)}}  \, , \qquad \quad 
 Q_m(\Sigma_{(2|0)}) = \int_{\SSM} F \wedge \mathbb{Y}^{(2|4)}_{\Sigma_{(2|0)}} \, .
\end{equation}
Using the current conservation laws, it is easy to see that these are topological operators, also invariant under super diffeomorphisms. 

The construction of the Poincar\'e dual  $\mathbb{Y}^{(2|0)}_{\Sigma_{(2|4)}}$ can be done using the immersion of $\Sigma_{(2|4)}$. Since this surface extends along two bosonic coordinates and four fermionic ones, the embedding is given by 
two functional independent equations $\Phi_1=0$ and $\Phi_2=0$. Consequently, we can write 
\begin{equation}
\label{PicA}
\mathbb{Y}^{(2|0)}_{\Sigma_{(2|4)}} = 
\delta(\Phi_1) d\Phi_1 \wedge 
 \delta(\Phi_2) d\Phi_2 \, . 
\end{equation}
 
Instead, immersion of the $\Sigma_{(2|0)}$ surface requires localizing two bosonic coordinates and all the fermionic ones. This is established by six equations, two bosonic $\Phi_i=0$, $i=1,2$, and four fermionic $\Theta_\a=0, \bar\Theta_{\dot\a} =0 $ with $\a, \dot\a =1,2$. The 
corresponding Poincar\'e dual reads
\begin{eqnarray}
    \label{ST_A}    \mathbb{Y}^{(2|4)}_{\Sigma_{(2|0)}} = 
    \delta(\Phi_1) d\Phi_1 \wedge 
 \delta(\Phi_2) d\Phi_2 \bigwedge_{\a=1}^2 \Theta_\a \delta( d \Theta_\a) \bigwedge_{\dot\a=1}^2 \bar\Theta_{\dot\a} \delta( d \bar\Theta_{\dot\a}) \, .
\end{eqnarray}

It is easy to verify that both operators \eqref{PicA} and \eqref{ST_A} are closed, but not exact \cite{Castellani:2015ata,Castellani:2023tip}. Changing the embedding functions amounts to adding $d$-exact terms to these expressions. Charges corresponding to closed currents are then invariant under these variations.  

We look for operators that are charged under the action of $Q_e(\Sigma_{(2|4)})$ and $Q_m(\Sigma_{(2|0)})$ defined in \eqref{eq:4dcharges}. 
Proceeding as described above, it is convenient to gauge the super-higher-form symmetries by turning on background gauge fields $B^{(2|0)}$ and $B_{CW}^{(2|2)}$ and compute 
\begin{equation}
\label{eq:4dpathint}
    Z[B^{(2|0)}, B_{CW}^{(2|2)}] = \int [{\mathcal D} A] e^{i(S+S_{min})} \, ,
\end{equation}
with 
\begin{equation}
    \label{eq:Smin4d}
    S+ S_{min} = \int_{\SSM} \left( - \frac12 (F - B^{(2|0)}) \wedge \star (F - B^{(2|0)}) +  F \wedge B_{CW}^{(2|4)} \right)
     \, .
\end{equation}
Under gauge transformations 
\begin{eqnarray}
\label{eq:transfs2}
    F \rightarrow F + d\Lambda^{(1|0)} 
    \, , \qquad  B^{(2|0)} \rightarrow B^{(2|0)} + d \Lambda^{(1|0)} \, , \qquad 
    B^{(2|4)}_{CW} \rightarrow B^{(2|4)}_{CW} + d \Lambda^{(1|4)} \, ,
\end{eqnarray}
the action is invariant up to an anomaly term. Precisely, 
\begin{equation}
\label{SGcwE3}
    S' + S_{min}' = S + S_{min} 
    + \int_{\SSM}  d\Lambda^{(1|0)} \wedge B^{(2|4)}_{CW} \, . 
\end{equation}

We first consider operators charged under the action of the symmetry generator $e^{i \alpha_e Q_e(\Sigma_{(2|4)})}$. Since this is the generator of a $(1|0)$-superform symmetry, we expect the charged operators to be $(1|0)$-dimensional objects, that is, super Wilson loops \cite{Cremonini:2020mrk}. According to the general structure \eqref{chA} specialized to $p=1$, these operators are defined as
\begin{eqnarray}
\label{STSL}
W({\mathcal C}) = e^{i q_e \Gamma_{\mathcal C}}\,, ~~~~~
\Gamma_{\mathcal C} = \int_{\SSM} A \wedge \mathbb{Y}^{(3|4)}_{\mathcal C} = \int_{\SSM} F \wedge \Theta_{\Omega_{(2|0)}}^{(2|4)} \, ,
\end{eqnarray}
where $A$ is the Maxwell $(1|0)$-superpotential, $q_e$ is the operator charge, and $\mathbb{Y}^{(3|4)}_{\mathcal C} = d \Theta_{\Omega_{(2|0)}}^{(2|4)}$ is the Poincar\'e  dual form of the line ${\mathcal C}$, which is the boundary of the $\Omega_{(2|0)}$ surface.

The path integral evaluating $\langle e^{i \alpha_e Q_e(\Sigma_{(2|4)})} W({\mathcal C}) \rangle$  corresponds to \eqref{eq:4dpathint} with background fields $B^{(2|4)}_{CW} = q_e \Theta_{\Omega_{(2|0)}}^{(2|4)}$ and $B^{(2|0)} = \a_e \mathbb{Y}^{(2|0)}_{\Sigma_{(2|4)}} $ turned on. Now, applying a gauge transformation \eqref{eq:transfs2} to get rid of $B^{(2|0)}$ and taking into account the anomaly term in \eqref{SGcwE3}, we eventually obtain 
\begin{eqnarray}
\label{STSK}
\Big\langle e^{i \alpha_e Q_e(\Sigma_{(2|4)})} W({\mathcal C}) \Big\rangle= e^{ i \alpha_e  q_e {\rm SLink}(\Sigma_{(2|4)}, {\mathcal C})} \Big\langle W({\mathcal C}) \Big\rangle \, ,
\end{eqnarray}
where the super-linking number can be alternatively written as 
\begin{eqnarray}
\label{STSM}
{\rm SLink}(\Sigma_{(2|4)}, {\mathcal C}) = 
\int_{\SSM} \Theta^{(2|4)}_{\Omega_{(2|0)}} \wedge \mathbb{Y}^{(2|0)}_{\Sigma_{(2|4)}} = \int_{\SSM} \mathbb{Y}^{(3|4)}_{\mathcal C} \wedge \Theta^{(1|0)}_{\Omega_{(3|4)}} \, ,
\end{eqnarray}
with $\mathbb{Y}^{(2|0)}_{\Sigma_{(2|4)}} = d  \Theta^{(1|0)}_{\Omega_{(3|4)}}$.
This is in agreement with the result found in 
\cite{Cremonini:2020zyn}.

We can proceed in a similar way to find the class of operators carrying non-trivial charge with respect to $e^{i \alpha_m Q_m(\Sigma_{(2|0)})} $, with $Q_m(\Sigma_{(2|0)})$ given in \eqref{eq:4dcharges}. As for the three-dimensional case, charged operators are supermonopoles, that is, supersurfaces $\Sigma_{(1|4)}$ of singularities which break the Bianchi identity as
\begin{equation}
\label{eq:singBI}
    dF = q_m \mathbb{Y}^{(3|0)}_{\Sigma_{(1|4)}} \, .
\end{equation}
Writing $\mathbb{Y}^{(3|0)}_{\Sigma_{(1|4)}}= d \Theta^{(2|0)}_{\Omega_{(2|4)}}$ for a non-compact surface ${\Omega_{(2|4)}}$ whose boundary is the supermonopole, the insertion of the supermonopole $\mathcal{D}({\Sigma_{(1|4)}})$ inside the path integral corresponds to turning on $B^{(2|0)}_{\mathcal{D}} = q_m \Theta^{(2|0)}_{\Omega_{(2|4)}}$. On the other hand, the inclusion of $e^{i \alpha_m Q_m(\Sigma_{(2|0)})} $ corresponds to turning on $B_{CW}^{(2|4)} = \a_m \mathbb{Y}^{(2|4)}_{\Sigma_{(2|0)}}$. Performing a gauge transformation \eqref{eq:transfs2} with $d \Lambda^{(1|4)} = - \a_m \mathbb{Y}^{(2|4)}_{\Sigma_{(2|0)}}$, we remove $B_{CW}^{(2|4)}$, but an anomaly arises due to the singularity in \eqref{eq:singBI} and we eventually obtain 
\begin{equation}
 \Big\langle e^{i \alpha_m Q_m(\Sigma_{(2|0)})}  \mathcal{D}({\Sigma_{(1|4)}}) \Big\rangle = e^{i \alpha_m  {\rm SLink}(\Sigma_{(2,0)}, {\Sigma_{(1|4)}})}  \Big\langle \mathcal{D}({\Sigma_{(1|4)}}) \Big\rangle    \, ,
\end{equation}
\label{STSOB}
where the super-linking number between the two supersurfaces is alternatively defined as
\begin{eqnarray}
\label{STSPB}
{\rm SLink}(\Sigma_{(2,0)}, \Sigma_{(1|4)}) = 
\int_{\SSM} \Theta^{(2|0)}_{\Omega_{(2|4)}} \wedge \mathbb{Y}^{(2|4)}_{\Sigma_{(2|0)}} = \int_{\SSM} \mathbb{Y}^{(3|0)}_{\Sigma_{(1|4)}} \wedge \Theta^{(1|4)}_{\Omega_{(3|0)}} \, , 
\end{eqnarray}
with $ \mathbb{Y}^{(2|4)}_{\Sigma_{(2|0)}}=  d\Theta^{(1|4)}_{\Omega_{(3|0)}}$.

\vspace{0.5cm}

\paragraph{Super-gCW currents.} In four dimensions, the only super-CW current of the form \eqref{superCW} is $J^{(4|0)} = F \wedge F$. However, its bosonic dimensions saturate the spacetime, thus leading to a degenerate case where the conservation law is trivial. This current can be interpreted as generating a $(-1|4)$-integral form symmetry. This type of current will be studied elsewhere \cite{grassi}. 

It is possible to enlarge the spectrum of super-CW symmetries by considering currents of the form $\omega \wedge F$, where $\omega$ are closed forms constructed in terms of the supervielbeins. Precisely, we consider
\begin{eqnarray}
    \label{cocA}
    \omega^{(3|0)} = V^{\alpha\dot\alpha} \wedge \psi_\alpha \wedge \bar \psi_{\dot\alpha}\,, ~~~~~
    \omega^{(4|0)} = (V\wedge V)^{\alpha\beta} \wedge \psi_\alpha \wedge \psi_{\beta}\,, ~~~~~
    \bar\omega^{(4|0)} = (V\wedge V)^{\dot\alpha\dot\beta} \wedge\bar\psi_{\dot\alpha} \wedge \bar\psi_{\dot\beta}\,, ~~~~~
\end{eqnarray}
They belong to the $H^{(\bullet|0)}(d)$ cohomology. Their closure easily follows from the identity $dV^{\a \dot\a} = \psi^\a \wedge \bar\psi^{\dot\a}$.
Analogously, the corresponding Hodge dual operators 
\begin{eqnarray}
    \label{cocB}
    &&\omega^{(1|4)} \equiv \star  \omega^{(3|0)} =  (V\wedge V\wedge V)^{\alpha\dot\alpha} \wedge \iota_\alpha \bar \iota_{\dot\alpha}  \delta^{(4)}(\psi)\,, ~~~~~ \nonumber \\
    &&\omega^{(0|4)} \equiv \star  \omega^{(4|0)} =(V\wedge V)^{\alpha\b} \wedge \iota_\alpha \iota_{\beta}  \delta^{(4)}(\psi)\,, ~~~~~ \nonumber \\
     &&\bar\omega^{(0|4)}  \equiv \star  \bar\omega^{(4|0)} =(V\wedge V)^{\dot\alpha\dot\b} \wedge \bar\iota_{\dot\alpha} \bar \iota_{\dot \beta} \delta^{(4)}(\psi)\,, 
\end{eqnarray}
are integral forms in the 
$H^{(\bullet|m)}(d)$ cohomology\footnote{Locally, for flat supermanifolds we have the following dualities     
\begin{equation}
\label{tenEA}
\star \omega^{(3|0)} \wedge \omega^{(3|0)} = 
\star \omega^{(4|0)} \wedge \omega^{(4|0)} =  \star \bar\omega^{(4|0)} \wedge \bar\omega^{(4|0)} =V^{6} \delta^{(8)}(\psi) = d^{6}x \; \delta^{8}(d\theta) 
\nonumber\, .
\end{equation}
}. 
Therefore, in the $D=(4|4)$ super-Maxwell theory, we can define three additional conserved integral form currents
\begin{equation}
\label{eq:omegacurrents}
    J^{(3|4)} = \omega^{(1|4)} \wedge F \, , \qquad 
    J^{(2|4)} = \omega^{(0|4)} \wedge F \, , \qquad 
    \bar{J}^{(2|4)} = \bar\omega^{(0|4)} \wedge F \, ,
\end{equation}
and the corresponding super-gCW topological operators 
\begin{eqnarray}
    \label{cocC}
    && Q(\Sigma_{(3|4)}) = \int_{\Sigma_{(3|4)}} \omega^{(1|4)} \wedge F = 
    \int_{\SSM}  \omega^{(1|4)} \wedge F \wedge \mathbb{Y}^{(1|0)}_{\Sigma_{(3|4)}}  \, , \nonumber \\
    && Q(\Sigma_{(2|4)}) = \int_{\Sigma_{(2|4)}} \omega^{(0|4)} \wedge F = 
    \int_{\SSM}  \omega^{(0|4)} \wedge F \wedge \mathbb{Y}^{(2|0)}_{\Sigma_{(2|4)}}  \, , \nonumber \\
    &&Q(\bar\Sigma_{(2|4)}) = \int_{\bar\Sigma_{(2|4)}} \bar\omega^{(0|4)} \wedge F = \int_{\SSM}  \bar\omega^{(0|4)} \wedge F \wedge \mathbb{Y}^{(2|0)}_{\bar\Sigma_{(2|4)}} \, .
\end{eqnarray}

In order to work out their explicit field dependence and physical meaning, one has to make a definite choice of the embedded supersurfaces and reduce them to ordinary Berezin integrals in superspace. 
We consider, for example, the first operator in \eqref{cocC}. The easiest way to parametrize the embedding of 
$\Sigma_{(3|4)}$ into $\SSM$ is by choosing 
\begin{eqnarray}
    \label{cocD}    \mathbb{Y}^{(1|0)}_{\Sigma_{(3|4)}} = V^{\a\dot\a} \chi_{\a\dot\a} + \psi^\alpha \eta_\alpha + \bar\psi_{\dot\alpha} 
    \bar\eta^{\dot \alpha} \, ,
\end{eqnarray}
where $\chi_{\a\dot\a},\eta_\alpha, \bar\eta_{\dot \alpha}$ are the moduli of the embedding, satisfying $d \chi_{\a\dot\a} = 0$, $d \eta_\alpha = -\frac12\chi_{\alpha \dot \alpha} \bar \psi^{\dot \alpha}$ and $d \bar\eta^{\dot\alpha} = \frac12\chi^{\alpha \dot \alpha} \psi_{\alpha}$, as required by the closure condition $d\mathbb{Y}^{(1|0)}_{\Sigma_{(3|4)}} =0$. We consider once again the explicit example of the $\mathbb{S}^{(3|4)}$ supersphere given by the defining equation \eqref{CA_BA}.
In terms of the supervielbeins, the corresponding Poincar\'e dual form $\mathbb{Y}^{(1|0)}_{\Sigma_{(3|4)}} = \Phi d \Phi$ is explicitly given by 
\begin{equation}
    \label{cocF}    \mathbb{Y}^{(1|0)}_{\Sigma_{(3|4)}} =
\delta\Big(x^{\alpha \dot\alpha} x_{\alpha \dot\alpha} + \mu\, (
    \theta^\alpha \theta_\alpha + 
    \bar\theta_{\dot\alpha} \bar\theta^{\dot\alpha}) - 1\Big)
    \Big(V^{\alpha\dot\alpha} x_{\alpha \dot\alpha} + 
    \psi^\alpha (\mu \, \theta_\alpha - \frac12 x_{\alpha\dot\alpha} \bar\theta^{\dot\alpha}) + 
     \bar\psi_{\dot\alpha} (\mu \,\bar\theta^{\dot\alpha} + \frac12 x^{\alpha\dot\alpha} \theta_{\alpha})\Big) \, .
\end{equation}
Comparing with the general expression \eqref{cocD}, we can infer the corresponding parameters $\chi_{\a\dot\a}, \eta_\alpha$ and $\bar\eta_{\dot\alpha}$. It is easy to verify that they satisfy the closure conditions.  

Inserting all the ingredients in the first integral in \eqref{cocC} (see equations  \eqref{STSA}, \eqref{cocB} and \eqref{cocF}), and taking into account that the only surviving terms are the ones proportional to $V^4 \delta^{(4)} (\psi) = [d^4x d^4 \theta]$, we eventually find
\begin{eqnarray}
    \label{cocE2}
     &&Q(\Sigma_{(3|4)}) 
 = \int_{\SM}  \omega^{(1|4)} \wedge F \wedge \mathbb{Y}^{(1|0)}_{\Sigma_{(3|4)}}   
     \\
     && 
     = \int (V\wedge V\wedge V)^{\alpha\dot\alpha} \iota_\alpha \bar \iota_{\dot\alpha}  \delta^{(4)}(\psi)  \wedge \Big(V^a V^b F_{ab} + \bar\psi_{\dot \b} V^{\b \dot\b} W_\b + 
   \bar W_{\dot \b} V^{\b \dot\b}  \psi_{\b} \Big) \wedge \Big( V^c \sigma_c + \psi^\gamma \eta_\gamma + \bar\psi_{\dot\gamma} 
    \bar\eta^{\dot \gamma} \Big) \nonumber \\
       && = \int  [d^4 x d^4 \theta] \, \delta\Big(x_{\alpha \dot\alpha} x^{\alpha \dot\alpha} + 
    \mu\, (\theta^\alpha \theta_\alpha + 
    \bar\theta_{\dot\alpha} \bar\theta^{\dot\alpha}) - 1\Big) 
    \Big(W^\alpha (\mu \theta_\alpha - \frac12  x_{\alpha\dot\alpha} \bar\theta^{\dot\alpha}) + \bar W_{\dot\alpha} (\mu \bar\theta^{\dot\alpha} + \frac12 x^{\alpha\dot\alpha} \theta_{\alpha})\Big) \, . \nonumber
\end{eqnarray}
The ordinary Berezin integral can be easily performed by expanding the delta function in powers of $\theta$'s producing spacetime derivatives acting on the integrand. 
Expanding the Dirac delta function, we eventually have 
\begin{eqnarray}
    \label{contoA}
     Q(\Sigma_{(3|4)}) &=& 
     \int  d^4 x  \, \delta\Big(x^2 - 1\Big) 
    \Big(- \mu \left. \bar D^2 D_\alpha W^\alpha\right|  + \frac12  x_{\alpha\dot\alpha} 
    \left. \bar D^{\dot \alpha} D^2 W^\alpha\right| + {\rm h.c.}
    \Big) \, . \nonumber \\
    &+&  \int  d^4 x  \, \delta'\Big(x^2 - 1\Big)
    (\mu^2 \left. D_\a W^\alpha \right| + {\rm h.c.}) \, ,
\end{eqnarray}
where the delta functions allow to perform the spacetime integral by localizing the result at $x^2 =1$. 

One can apply a similar reasoning for the other two charges in \eqref{cocC}. Since nothing deviates significantly from this calculation, except for a change in the definition \eqref{cocF} of the Poincar\'e dual, we are not going to report the results. 

Looking for operators charged under the action of gCW symmetries generated by \eqref{cocC}, we now prove that the supermonopole $\mathcal{D}(\Sigma_{(1|4)})$ introduced above is also charged respect to them. To this end, we first couple the gCW currents \eqref{eq:omegacurrents} to background gauge fields, which amounts to considering in the path integral the generalized action
\begin{equation}
    S + S_{min} = \int_{\SSM} \left( - \frac12 (F - B^{(2|0)}) \wedge \star (F - B^{(2|0)}) +  J^{(3|4)} \wedge B^{(1|0)} + J^{(2|4)} \wedge B^{(2|0)} + \bar{J}^{(2|4)} \wedge \bar{B}^{(2|0)} \right) \, .
\end{equation}
As long as the gCW currents are conserved, this action is invariant under the following background gauge transformations 
\begin{equation}
\label{eq:CWgauge transfs}
    B^{(1|0)} \to B^{(1|0)} + d\Lambda^{(0|0)} \, , \qquad 
    B^{(2|0)} \to B^{(2|0)} + d\Lambda^{(1|0)} \, , \qquad 
    \bar{B}^{(2|0)} \to \bar{B}^{(2|0)} + d\bar\Lambda^{(1|0)} \, .
\end{equation}

To be definite, we consider the action of the symmetry generator $U(\Sigma_{(3|4)},\a) = e^{i \a Q(\Sigma_{(3|4)})}$ on the supermonopole. 
Given the explicit expression of the charge in \eqref{cocC}, the insertion of this operator inside the path integral corresponds to turning on $B^{(1|0)} = \a \mathbb{Y}^{(1|0)}_{\Sigma_{(3|4)}}$. Therefore, we can write
\begin{equation}
    \Big\langle U(\Sigma_{(3|4)},\a) \, \mathcal{D}(\Sigma_{(1|4)}) \Big\rangle = Z[B_{\mathcal{D}} = q_m \Theta^{(2|0)}_{\Omega_{(2|4)}}, B^{(1|0)} = \a \mathbb{Y}^{(1|0)}_{\Sigma_{(3|4)}}, 0, 0] \, .
\end{equation}
We should now perform a $\Lambda^{(0|0)}$-transformation \eqref{eq:CWgauge transfs} to get rid of the $U$ generator. However, since the supermonopole breaks the $F$ conservation law - see equation  \eqref{eq:singBI} - the $J^{(3|4)}$ current is no longer conserved, rather
\begin{equation}
    d J^{(3|4)} = q_m \, \omega^{(1|4)} \wedge \mathbb{Y}^{(3|0)}_{\Sigma_{(1|4)}} \, .
\end{equation}
Therefore, under a gauge transformation of the action, the residual term $\int dJ^{(3|4)} \wedge \Lambda^{(0|0)}$ gives rise to a non-vanishing anomaly. Inserting the explicit expressions for the background fields and the gauge parameter, we can eventually write 
\begin{equation}
\label{eq:gCWcharge0}
    \Big\langle U(\Sigma_{(3|4)},\a) \, \mathcal{D}(\Sigma_{(1|4)}) \Big\rangle  = e^{-i \a \, q_m \int_{\SSM} \Theta^{(2|0)}_{\Omega_{(2|4)}} \wedge \, \omega^{(1|4)} \wedge \mathbb{Y}^{(1|0)}_{\Sigma_{(3|4)}} }
\Big\langle \mathcal{D}(\Sigma_{(1|4)}) \Big\rangle \, ,
\end{equation} 
where we recall that $\Omega_{(2|4)}$ is the supersurface whose boundary is the 't Hooft supersurface $\Sigma_{(1|4)}$ defining the supermonopole. The extra phase, which arises from the anomalous gauge transformation, can be interpreted as the superlinking number between $\Sigma_{(1|4)}$ and a $(2|0)$-supersurface defined by the Poincar\'e dual 
\begin{equation}
\widetilde{\mathbb{Y}}^{(2|4)}_{\tilde\Sigma_{(2|0)}} \equiv \omega^{(1|4)} \wedge \mathbb{Y}^{(1|0)}_{\Sigma_{(3|4)}}  \, .
\end{equation}
This operator is indeed closed but not exact; therefore, it defines a genuine Poincar\'e dual. In order to check that it is non-trivial, we consider the expression in \eqref{cocF} 
and multiply it by $\omega^{(1|4)}$ given in 
\eqref{cocB}. Using definition \eqref{CA_BA} the result is
\begin{eqnarray}
    \label{newcocA} \widetilde{\mathbb{Y}}^{(2|4)}_{\tilde\Sigma_{(2|0)}}   &=& (V\wedge V\wedge V)^{\alpha\dot\alpha} \wedge \iota_\alpha \bar \iota_{\dot\alpha}  \delta^{(4)}(\psi) \wedge \delta(\Phi) d \Phi  \\
    &=& \delta(\Phi) \left( V^4\, x^{\a\dot\a} \iota_\alpha \bar \iota_{\dot \a} + (V^3)^{\alpha\dot\alpha}
    (\mu \theta_\alpha - \frac12 x_{\a\dot\b} \bar \theta^{\dot\b}) \bar\iota_{\dot\alpha} + 
    (V^3)^{\alpha\dot\alpha} 
    (\mu \bar\theta_{\dot\alpha} - \frac12 x_{\b\dot\a}  \theta^{\b})\iota_{\alpha}
   \right)  \delta^{(4)}(\psi)  \, . \nonumber
\end{eqnarray}
All the pieces are non-trivial. Plugging them into a supermanifold integral like the one in \eqref{eq:gCWcharge0}, we generally obtain a non-vanishing result that can be eventually expressed as an ordinary superspace integral. 

The $\tilde\Sigma_{(2|0)}$ supersurface can be seen as a $(1|4)$-foliation of $\Sigma_{(3|4)}$.
It is not difficult to see that the superlinking number is non-trivial as long as $\tilde\Sigma_{(2|0)}$ and $\Omega_{(2|4)}$ intersect; in other words, if $\tilde\Sigma_{(2|0)}$ and the 't Hooft supersurface are linked. 

The same procedure can be easily applied to the other two symmetries generated by $Q(\Sigma_{(2|4)})$ and $Q(\bar\Sigma_{(2|4)})$ in \eqref{cocC}. The $\mathcal{D}(\Sigma_{(1|4)})$ supermonopole carries a non-trivial charge also with respect to these two symmetries, given by the superlinking number between the 't Hooft supersurface and a $(0|4)$-foliation of the supersurfaces supporting the charge generators.

 \subsection{$D=(6|8)$}

 We now consider $N=(1,0)$ super-Maxwell theory in six dimensions defined in $D=(6|8)$ superspace \cite{Siegel:1978yi, Howe:1983fr}. This is conveniently described by a set of coordinates $(x^a, \theta_\a^A)$, where $x^a$, $a=0, \dots, 5$ are the bosonic directions. At the same time, $\theta_\a^A$ are a set of symplectic Majorana-Weyl spinors, that is, four-components complex spinors carrying a  $SU(2)$ index $A=1,2$, a $SU(4)$ spinorial representation index $\alpha =1, \dots,4$, and satisfying the $SU(2)$-Majorana condition, $(\theta_\a^A)^* \equiv \bar{\theta}_{\dot\a \, A} =\epsilon_{AB} {\mathcal B}_{\dot\a}^{\; \b} \theta_\b^B$ for a unitary matrix ${\mathcal B}$ with  ${\mathcal B}^* {\mathcal B} = -1$. In this superspace, there is no rising-lowering matrix for spinorial indices. We use $4 \times 4$  Pauli-Dirac gamma matrices $\delta_\alpha^\beta, \gamma^a_{\alpha\beta}, \gamma^{a,\alpha\beta}$, and a set of supercovariant derivatives satisfying $\{ D^{A \, \a}, D^{B \, \b} \} = i \epsilon^{AB} \gamma^{a , \a\b} \partial_a \equiv i \epsilon^{AB} \partial^{\a\b}$. Supervielbeins $V_{\a\b} = dx_{\a\b} + \epsilon_{AB} \theta_\a^A d\theta_\b^B $ and $\psi^A_\alpha = d\theta_\a^A$ 
 satisfy the Maurer-Cartan equations 
 \begin{eqnarray}
     \label{sixE}
     d V_{\a\b}  = \epsilon_{AB}  \, \psi^A_\a \wedge \psi_\b^B \,  \,, ~~~~~  d \psi^A_\alpha =0\, ,  
 \end{eqnarray}
with $d = V_{\a\b} \, \partial^{\a\b} +  \epsilon_{AB} \, \psi^A_\a \, D^{B \, \a}  $.
 
 The super-Maxwell multiplet consists of a vector superfield $A_{[\a\b]}$, a Majorana-Weyl spinor $W^A_\alpha$ and a triplet of auxiliary superfields $D_{(AB)}$. Imposing the conventional constraint that removes the $\psi \wedge \psi$ term, the expansion of the $(2|0)$-superform field strength reads
\begin{equation}
     \label{sixA}
     F = F^{\a\b,\g\d} \, V_{\a\b} \wedge V_{\g\d} 
     + \epsilon_{AB} \epsilon^{\a\b\g\d} W^A_\alpha \, \psi^B_\b \wedge V_{\g\d}\, , 
\end{equation}
with $F^{\a\b,\g\d} = - F^{\g\d, \a\b}$ and 
$F^{\a\b,\g\d} = - F^{\b\a, \g\d}$, which has 15 independent components. 
     Bianchi identities lead to the following equations 
     \cite{Howe:1983fr}
 \begin{eqnarray}
     \label{sixB}
     \partial^{[\a\b} F^{\g\d ,\r\s]} &=&0 \,, \nonumber \\
     D^\a_A F^{\b\g ,\r\s} &=& \epsilon_{AB} \, \epsilon^{\eta\a [\r\s} \partial^{\b\g]} W_\eta^B \,, \nonumber \\
     3 D^{A\alpha} W^B_{\beta} &=& \epsilon^{AB} \epsilon_{\b\g\d\r}F^{\a\g, \d\r} + \delta^\alpha_{~\beta} D^{(AB)} \, .
 \end{eqnarray}
It relates the gaugino field strength $W^A_\alpha$ to gauge field strength $F^{\a \b,\g\d}$ and to the auxiliary field triplet $D^{(AB)}$. Note that the Bianchi identities do not imply the equations of motion; therefore, this is an off-shell multiplet. 

To write the super-Maxwell action in \eqref{eq:superaction} and provide the corresponding equations of motion, we need to define the Hodge dual of $F$. However, this cannot be consistently defined due to the lack of a lowering-rising Dirac matrix for spinorial indices in flat superspace. The problem can be circumvented by introducing a constant non-trivial flux background for the three-form $H_{abc}$ in $D=6, N=1$ supergravity \cite{Nahm:1977tg,Louis:2016tnz,Karndumri:2016ruc,DAuria:2000afl,Andrianopoli:2001rs}, such that $C_{\alpha\beta} =(\gamma^{abc} H_{abc})_{\alpha\beta}$ is a constant symmetric matrix in spinorial space. We can use it to build up the supermetric needed to define the Hodge dual. At the end of the computation, we send the flux to zero. 
Armed with this definition of Hodge dual, we define 
\begin{eqnarray}
    \label{sixC}
    \star F = F^{\a\b,\g\d} (V\wedge V \wedge V\wedge V)_{\a\b, \g\d} \, \delta^{(8)}(\psi) + 
    W^A_\a  C_{\b\g}\iota^\b_A (V\wedge V \wedge V\wedge V \wedge V )^{\a\b}\delta^{(8)}(\psi) \, , 
\end{eqnarray}
where we have defined $(V\wedge V \wedge V\wedge V)_{\a\b, \g\d} \equiv (\g^a)_{\a\b} (\g^b)_{\g\d} \epsilon_{abcdef} V^c \wedge V^d \wedge V^e \wedge V^f$ and $(V\wedge V \wedge V\wedge V \wedge V)^{\a\b} \equiv (\g^a)^{\a\b} \epsilon_{abcdef} V^b \wedge V^c \wedge V^d \wedge V^e \wedge V^f$. 
Imposing the closure $d \star F=0$ leads to the following equations of motion
\begin{eqnarray}
    \label{sixD}
    \epsilon_{\a\b\g\d}\partial^{\a\b} F^{\g\d,\r\s}  + \epsilon^{\r\s\b\d} D^\a_A C _{\a\b} W^A_\d = 0\,, ~~~~~
    \partial^{\a\b} W^A_\b =0\,. 
\end{eqnarray}
 We retrieve the super-Maxwell equations by taking the  $C_{\alpha\beta} \rightarrow 0$ limit.

 \vspace{0.3cm}

\paragraph{Electric and magnetic currents.} As in the previous cases, the $N=(1|0)$ super-Maxwell theory in six dimensions enjoys a $(1|0)$-superform symmetry generated by $\star F$ and a topological $(3|8)$-integral form symmetry generated by $F$. Operators charged under these symmetries can be easily constructed. They are a super Wilson loop of the form \eqref{chA} with $p=1$, and a supermonopole supported by a $\Sigma_{(3|8)}$ supersurface of singularities for the $F$ current, which is then no longer conserved everywhere in the supermanifold, in analogy with what happens in the bosonic theory (see eq.  \eqref{eq:monopole}). Details of the construction are given in appendix \ref{sect:CWcurrents}. 

\paragraph{Super-(g)CW currents.} In $D=(6|8)$ superspace, we have enough dimensions for constructing further higher-form conserved currents. First of all, we can define the generalized Chern-Weil supercurrent $J^{(4|0)} = \frac12 F \wedge F$, which generates a $(1|8)$-integral form symmetry. The corresponding charged objects are operators supported on  $(1|8)$-dimensional supersurfaces obtained by intersecting two supermonopole 't Hooft surfaces. The construction is detailed in appendix 
\ref{sect:CWcurrents}. 

In addition, we can look for closed currents of the form $\omega \wedge J$, where $\omega$ are closed forms constructed in terms of the supervielbeins. In this case, there is an element of the cohomology, whose representative can be 
written as 
\begin{eqnarray}
    \label{sixF}
    \omega^{(3|0)} = V^{\a\b} \wedge \psi^A_\a  \wedge \psi^B_\b \, \epsilon_{AB}\, .
\end{eqnarray}
It is closed due to the Fierz identities and the Schoutens identities for 
$SU(2)$ indices. 
The 3-form $\omega^{(3|0)}$ is an element of the Chevalley-Eilenberg cohomology (see for example \cite{Cremonini:2022cdm,Cremonini:2024ddc}). A strategy known as Free Differential Algebras (FDA) allows the reconstruction of the full supergravity spectrum by replacing $\omega^{(3|0)}$ with the $d$-variation potential $B^{(2|0)}$. Together with the graviton and the gravitinos $(V^{\alpha\beta}, \psi^A_\alpha)$ this set of fields form the spectrum of $D=6, N=1$ supergravity 
(see for example \cite{Castellani:1991et}).

The Hodge dual of $ \omega^{(3|0)} $ is easily computed \cite{Castellani:2015ata}
\begin{eqnarray}
    \label{sixG}
    \star \omega^{(3|0)} \equiv \omega^{(3|8)} = (V\wedge V  \wedge V \wedge V \wedge V)_{\a\b}  \wedge \iota^\a_A  \iota_B^\b \epsilon^{AB} \delta^{(8)}(\psi) \, \, ,
\end{eqnarray}
where $\iota_A^\alpha$  is the derivative with respect to $\psi^A_\alpha$. Using these closed forms, we can construct the following new superform and integral form currents
\begin{equation}
 J^{(5|0)} = \omega^{(3|0)} \wedge F \, , \qquad \quad 
 J^{(5|8)} = \omega^{(3|8)} \wedge F \, ,
\end{equation}
which generate a $(0|8)$-integral form symmetry and a $(0|0)$-superform symmetry, respectively. As in the four-dimensional case, the supermonopole is also charged under these gCW symmetries. We prove this statement in appendix \ref{sect:CWcurrents} and provide the corresponding charges.  

When we take the flat limit, the currents constructed with the $\omega$-forms have an interesting interpretation as remnants of supergravity couplings. 
In fact, in flat superspace  $\omega^{(3|0)}$ and $\omega^{(3|8)}$ are closed superforms, therefore locally they can be written as 
\begin{eqnarray}
    \label{SGA0}
    \omega^{(3|0)} = d B^{(2|0)} \,, ~~~~~~~
    \omega^{(3|8)} = d B^{(2|8)} \,, ~~~~~~~
\end{eqnarray}
where, as already mentioned, $B^{(2|0)}$ can be interpreted as the conventional two-form potential of $N=1$ supergravity multiplet. The additional potential $B^{(2|8)}$ is defined in terms of the dual, 
$\star d  B^{(2|0)} = d B^{(2|8)}$, and it does not introduce any new degrees of freedom. 

If we consider super-Maxwell theory coupled to dynamical $N=1$ supergravity \cite{DallAgata:1997yqq}, non-trivial curvatures arise, which modify  the first identity in \eqref{SGA0} as
\begin{eqnarray}
    \label{SGAA0}
    H^{(3|0)} = d B^{(2|0)} -  \omega^{(3|0)}\, . 
\end{eqnarray}
Exploiting the closure of $F$, the charge generator associated to $J^{(5|0)}$ can then be written as
\begin{eqnarray}
\label{SGB0}
Q(\Sigma_{(5|0)} ) = \int_{\Sigma_{(5|0)}}
\omega^{(3|0)}\wedge  F  
= 
-\int_{\Sigma_{(5|0)}}
H^{(3|0)}\wedge  F \,.  
\end{eqnarray}
The $H^{(3|0)}$ supergravity field strength plays the role of a background gauge field that gauges the $F$ symmetry. Since in a curved spacetime $dH^{(3|0)} \sim {\rm tr} R^2 \neq 0$, this charge is no longer topological and cannot be interpreted as being associated with a global symmetry. The Chern-Weil global symmetry is restored in the flat limit. This is consistent with the property that no exact global symmetries are allowed in quantum gravity \cite{Harlow:2018jwu,Harlow:2018tng} (see also \cite{Heidenreich:2020pkc} for a similar discussion).

\subsection{$D=(10|16)$}

As a final example of super-Maxwell theory, we consider the case of $N=1$ super-Maxwell theory in ten dimensions. In $D=(10|16)$ chiral superspace, we use the same approach as above and write \cite{Fre:2017bwd}
\begin{eqnarray}
\label{sA}
 F = V^{\a\b} \wedge V^{\g\d} F_{\a\b,\g\d} +  V_{\a\b} \wedge \psi^\a \,  W^\b  \, ,
 \end{eqnarray}
 where  $F_{ab} \equiv \gamma_a^{\a\b} \gamma_b^{\g\d} F_{\a\b,\g\d}$ and $W^\alpha$ are the field strength superfields and the gluino field, respectively\footnote{We applied already the conventional constraint $F_{\a\b}=0$. Ten dimensional Pauli-Dirac matrices $\gamma^a$ are taken real and symmetric \cite{Brink:1976bc, Harnad:1985bc}.}.

From the Bianchi identity $dF=0$, we obtain the following set of constraints for the field components
\begin{eqnarray}
\label{STSB2}
\partial_{[a} F_{bc]} = 0 \, , ~~~~~ D_\alpha W^\alpha =0\,,~~~~~ F_{ab} = D^\a (\gamma_{ab})_{\a\b} W^\b \,, ~~~~~
D_\alpha F_{ab} = (\gamma_{[a})_{ \alpha\beta} \partial_{b]}W^\beta \, .
\end{eqnarray}

In ten dimensions, Bianchi identities automatically imply the equations of motion 
\begin{eqnarray}
\label{sE}
\partial^{a} F_{ab}=0\,,~~~~ \gamma^a \partial_{a} W_\beta =0 \,, 
\end{eqnarray}
These can be rephrased in the usual form $d \star F = 0$, provided that we introduce a suitable definition of Hodge dual, which in turn requires introducing a rising-lowering Dirac matrix $C_{[\alpha\beta]}$. This can be done in a non-trivial background. For example, if a 3-form $H$ is switched on, such for example in a non-trivial string background, one can take 
$C_{[\alpha\beta]} = (\gamma^{abc} H_{abc})_{\alpha\beta}$. Assigned this antisymmetric tensor, the Hodge dual of $F$ has the following structure
\begin{eqnarray}
    \label{sEA}
\star F = F_{ab} (V\wedge \dots \wedge V)^{[ab]} \delta^{(8)}(\psi) + 
(W \gamma_a C \iota)(V\wedge \dots \wedge V)^{a}  \delta^{(8)}(\psi) \, ,
\end{eqnarray}
where $\iota_\alpha$ is the derivative w.r.t. $\psi^\alpha$. Imposing the closure of $\star F$, we obtain
\begin{eqnarray}
    \label{sEB}
  \partial^a F_{ab} + D C \gamma_b W=0\,,~~~~ \gamma^a \partial_{a} W_\beta =0 \,,   
\end{eqnarray}
which reduce correctly to \eqref{sE} in the limit $C\rightarrow 0$. 

We now discuss higher-superform $U(1)$ symmetries. Since the technical details of the construction are very similar to the ones given in appendix \ref{sect:CWcurrents} for the $D=(6|8)$ case, we limit ourselves to listing the main results.   

\paragraph{Super-CW currents.} For the super-Maxwell theory in ten dimensions, beyond the electric current $J^{(8|16)} = \star F$, we have a set of CW currents of the form \eqref{superCW}. Precisely, 
\begin{eqnarray}
 \label{tenC}
 Q_k (\Sigma_{(2k|0)}) =  \int_{\Sigma_{(2k|0)}} J^{(2k|0)} =  \int_{\SSM} J^{(2k|0)} \wedge 
\mathbb{Y}^{(10 - 2k| 16)}_{\Sigma_{(2k|0)}} \,, \qquad k = 1,2,3,4,5 ~~~~~
\end{eqnarray}
is a set of topological operators which gives rise to super-higher-form symmetries of $(9-2k|16)$-type. Considering the magnetic charge $Q_m \equiv Q_1$, the corresponding charged supermonopole can be constructed as in the previous examples. In this case, it is given by a $(7|16)$-dimensional 't Hooft supersurface embedded in the supermanifold and corresponds to turning on the background 
$B_{\mathcal{D}} = q_m \Theta^{(2|0)}_{\Omega_{(8|16)}}$ in the path integral, with $\partial \Omega_{(8|16)} = \Sigma_{(7|16)}$. Its magnetic charge can be evaluated by following the same steps as in the lower dimensional cases, and it is given by the super-linking number between the supersurface supporting the symmetry generator and the 't Hooft singular supersurface associated with the supermonopole.  

Analogously, operators charged under $Q_k (\Sigma_{(2k|0)})$ for $k>1$ are generalized supermonopoles, whose 't Hooft supersurface is given by a $(9-2k|16)$-dimensional intersection of $k$ $\Sigma_{(7|16)}$
monopole supersurfaces. Details follow closely what has been done in appendix \ref{sect:CWcurrents} for the $D=(6|8)$ case.

\paragraph{Super-gCW currents.}
We can also construct gCW closed currents in ten dimensions. To this end, we introduce two geometric invariant forms 
\begin{eqnarray}
\label{tenD}
\omega^{(3|0)} = \psi \wedge \gamma^a \psi \wedge V_a\,,~~~~~~
\omega^{(7|0)} = \frac{1}{5!}\psi \wedge  \gamma^{a_1 \dots a_5} \wedge \psi \wedge V_{a_1} \wedge \dots \wedge V_{a_5}\,,~~~~~~
\end{eqnarray}
which belong to the cohomology $H^{(\bullet|0)}(d)$, thanks to Fiersz identities.  
Correspondingly, the Hodge duals are given by
\begin{eqnarray}
\label{tenE}
\star \omega^{(3|0)} &\equiv& \omega^{(7|16)} = \frac{1}{9!}\epsilon_{a_1 \dots a_9 b} V_{a_1} \wedge \dots \wedge V_{a_9}  \wedge 
\iota \gamma^b \iota \delta^{(16)}(\psi) \,, \nonumber \\
\star \omega^{(7|0)} &\equiv& \omega^{(3|16)} = \frac{1}{5!}\epsilon_{a_1 \dots a_5 b_1 \dots b_5} V_{a_1} \wedge \dots \wedge V_{a_5} \wedge 
\iota \gamma^{b_1 \dots b_5}  \iota \delta^{(16)}(\psi) \,. 
\end{eqnarray}
These are closed but not exact integral forms and belong to the $H^{(\bullet|16)}(d)$ cohomology. 

Now, in terms of these geometric forms, and using the CW currents \eqref{superCW}, we can build the following new  conserved currents 
\begin{eqnarray}
\label{tenF}
&&J^{(2k+3|0)} = \omega^{(3|0)}\wedge  J^{(2k|0)}\,, ~~~~~~~
J^{(2k+3|16)} = \omega^{(3|16)}\wedge  J^{(2k|0)}\,, ~~~~~~~ k = 1,2,3 \nonumber \\
&&J^{(9|0)} = \omega^{(7|0)} \wedge F \,, ~~~~~~~
J^{(9|16)} = \omega^{(7|16)} \wedge F\,, ~~~~~~~ \nonumber \\
&&J^{(8|16)} = \omega^{(3|0)} \wedge \omega^{(3|16)} \wedge F \, .
\end{eqnarray}

As a non-trivial case, we elaborate on one of these new currents, that is 
\begin{eqnarray}
    \label{exA}
J^{(9|16)} = 
    \omega^{(3|16)}\wedge  F \wedge F \wedge F \, ,
\end{eqnarray}
supported on a $\mathbb{S}^{(9|16)}$ supersphere, whose defining equation in the supermanifold is
\begin{eqnarray}
    \label{tenHA}
    x_a x^a + \theta \gamma_{abc}\theta \, \theta \gamma^{abc}\theta = 1 \, .
\end{eqnarray}
The corresponding Poincar\'e dual form is given by 
\begin{eqnarray}
    \label{tenHB}
    \mathbb{Y}^{(1|0)}_{\mathbb{S}^{(9|16)}} = 
    2 \delta(  x_a x^a + \theta \gamma_{abc}\theta \, \theta \gamma^{abc}\theta - 1)
     \Big(V^a x_a + \psi \Big(\frac12 x^a(\gamma_a \theta) +\gamma_{abc} \theta \, \theta\gamma_{abc} \theta\Big)\Big) \,. 
\end{eqnarray}
We are then ready to compute the associated generator. Inserting the explicit expressions \eqref{tenE} and \eqref{tenHB}, and using the distributional law \- $\iota_\alpha \delta^{(16)}(\psi) \psi^\beta = - \delta_{\alpha}^{~\beta}  \delta^{(16)}(\psi)$ together with some properties of Dirac matrices, we eventually find 
\begin{eqnarray}
\label{exB}
&&
\hspace{-1cm} Q = \int J^{(9|16)} \wedge  \mathbb{Y}^{(1|0)}_{\mathbb{S}^{(9|16)}} =   \\
&=&
\int \frac{1}{5!}\epsilon_{a_1 \dots a_5 b_1 \dots b_5} V^{a_1} \wedge \dots \wedge V^{a_5}  \iota \gamma^{b_1 \dots b_5}  \iota \delta^{(16)}(\psi) \nonumber \\
&&  \wedge 
\Big(V^a V^b F_{ab} + \psi \gamma^a W V_{a}\Big )\wedge \Big(V^r V^s F_{rs} + \psi \gamma^b W V_{b} \Big) \wedge \Big(V^a V^b F_{ab} + \psi \gamma^a W V_{a}\Big )\nonumber \\
&& \wedge 
 2 \delta(  x_a x^a + \theta \gamma_{abc}\theta \theta \gamma^{abc}\theta - 1)
     \Big(V^a x_a + \psi \Big(\frac12 x^a(\gamma_a \theta) +\gamma_{abc} \theta \theta\gamma_{abc} \theta\Big)\Big)
\nonumber \\
&=& 
\int  \delta(  x_a x^a + \theta \gamma_{abc}\theta \, \theta \gamma^{abc}\theta - 1) 
\left( c_1 F_{ab} W \gamma^{abc} W x_c + c_2 F_{ab} F_{cd} W \gamma^{abcd}\gamma^e\theta x_e \right) d^{10}x \, d^{16} \theta \, ,
\nonumber 
\end{eqnarray} 
where the coefficients $c_1$ and $c_2$ are numerical factors coming from the algebraic manipulations of Dirac matrices and contractions. Their values are irrelevant to our purposes. 
By expanding the delta function in powers of $\theta$'s, one can now perform the Berezin integration, obtaining an expression given in terms of field components.
Note that the result in \eqref{exB} is cubic in the quantum fields as a consequence of the structure of the original current. Therefore, several field components will get generated under the $\theta$-integration.  

Regarding the construction of operators carrying non-trivial charge under these super-gCW symmetries, in analogy with lower dimensional cases (see appendix \ref{sect:CWcurrents}), we expect the supermonopole charged with respect to the magnetic generator $Q_m$ to be charged also respect to the symmetries whose currents \eqref{tenF} contain a single $F$ factor. These are $J^{(5|0)}, J^{(5|16)}, J^{(9|0)}, J^{(9|16)}$ and $J^{(8|16)}$. Analogously, operators charged under symmetries generated by the currents $J^{(2k+3|0)}, J^{(2k+3|16)}$ in \eqref{tenF} with $k=2,3$ will correspond to generalized supermonopoles supported on singular surfaces given by a suitable intersection of $k$ supermonopole 't Hooft surfaces.  

\vskip 10pt

The geometric-CW supercurrents can be interpreted as coming from the flat limit of super-Maxwell coupled to dynamical $N=1$ supergravity.

To begin with, since $\omega^{(3|0)}$ and $\omega^{(7|0)}$ are cohomology classes they can be locally written as
\begin{eqnarray}
    \label{SGA}
    d B^{(2|0)} = \omega^{(3|0)}\,, ~~~~~~~
    d B^{(6|0)} = \omega^{(7|0)}\,, ~~~~~~~
\end{eqnarray}
where $B^{(2|0)}$ is the conventional Kalb-Ramond of type I supergravity. The additional potential $B^{(6|0)}$ is the dual 
$\star d  B^{(2|0)} = d B^{(6|0)}$ and it does not introduce new degrees of freedom. 
In this way, we reconstruct the complete supergravity spectrum. 

Making supergravity dynamical implies that we have to replace \eqref{SGA} with non-trivial fluxes
\begin{eqnarray}
    \label{SGAA}
    H^{(3|0)} = d B^{(2|0)} -  \omega^{(3|0)}\,, ~~~~~~~
    H^{(7|0)} = d B^{(6|0)} -  \omega^{(7|0)}\, . ~~~~~~~
\end{eqnarray}
It follows that the charges associated with the 
gCW currents in \eqref{tenF} can be written as 
\begin{eqnarray}
\label{SGB}
\int_{\Sigma_{(2k+3|0)}}
 J^{(2k+3|0)}
&=&
\int_{\Sigma_{(2k+3|0)}}
H^{(3|0)}\wedge  J^{(2k|0)}\,, \nonumber \\
\int_{\Sigma_{(9|0)}}
J^{(9|0)} &=&
\int_{\Sigma_{(9|0)}}
H^{(7|0)}\wedge  F\,, \nonumber \\
\int_{\Sigma_{(8|16)}}
J^{(8|16)}
&=&
\int_{\Sigma_{(8|16)}}
H^{(3|0)}\wedge \omega^{(3|16)} \wedge F \, .
\end{eqnarray}
As for the six-dimensional case, if the theory is coupled to dynamical supergravity, $H^{(3|0)}$ and $H^{(7|0)}$ are not closed, and these operators are no longer topological objects. As a consequence, they are not genuine generators of global higher-superform symmetries. These are recovered only in the flat limit, consistently with the fact that there should be no global symmetries in supergravity. This discussion clarifies the supergravity origin of the super-gCW symmetries.

\section{Super-Chern-Simons theory in $D=(3|2)$}
\label{sec:CS}
It is interesting to see how the construction of super-higher-form symmetries works in the three-dimensional $N=1$ Chern-Simons theory. 

Assigning an abelian $(1|0)$-superform potential $A$ in the $D=(3|2)$ supermanifold, the super-Chern-Simons action reads \cite{Cremonini:2019aao} \begin{equation}
\label{eq:CSaction}
S = \int_{\SSM} (A \wedge dA + \frac 12 V^3 \, W^\alpha W_\alpha) \wedge \mathbb{Y}^{(0|2)} \, ,
\end{equation}
where the Poincar\'e dual $\mathbb{Y}^{(0|2)}$ describes the immersion of the three-dimensional bosonic submanifold into $\SSM$. It can be taken as the Hodge dual of the cohomology element
  $\omega^{(3|0)} = \psi \wedge \gamma_a \psi \wedge V^a$, precisely
  \begin{eqnarray}
\label{SCS}
\mathbb{Y}^{(0|2)} \equiv \star \omega^{(3|0)}  = \epsilon_{abc} V^a \wedge V^b \wedge \iota \gamma^c \iota \delta^2(\psi) \, .
\end{eqnarray}
It is easy to see that by inserting this expression in \eqref{eq:CSaction} together with the $A$ expansion, we reproduce the well-known Chern-Simons action in superspace \cite{Nishino:1991sr}. 

The equations of motion  $F = dA = 0$ constrain the theory to have no propagating degrees of freedom. In expansion \eqref{S3A}, this implies 
the vanishing of the field strength $F_{ab}$ (up to a pure gauge term) and the gravitino, $W^\alpha=0$. 

In this case, the current $J^{(1|0)}=A$ is on-shell closed and gives rise to a $(1|2)$-integral form symmetry generated by 
\begin{eqnarray}
\label{SCA}
U(\mathcal{C},\a) = e^{i\a Q(\mathcal{C})} \, , \quad {\rm with} \qquad Q(\mathcal{C}) = \int_\mathcal{C} A = \int_{\SSM} A\wedge \mathbb{Y}^{(2|2)}_\mathcal{C}  \, .
\end{eqnarray}
This is nothing but the super Wilson line in $D=(3|2)$. 

Operators charged under this symmetry are $(1|2)$-dimensional objects of the form \eqref{chA2}. Sticking to that notation, they read
\begin{equation}
    \tilde{W}(\Sigma_{(1|2)}) = 
    e^{i\tilde{q} \int_{\Sigma_{(1|2)}} A \wedge \mathbb{Y}^{(0|2)}} = 
    e^{i \tilde{q}\int_{\SSM} A \wedge \mathbb{Y}^{(0|2)} \wedge \mathbb{Y}^{(2|0)}_{\Sigma_{(1|2)}}} \, , 
\end{equation}
with $\mathbb{Y}^{(0|2)}$ given in \eqref{SCS}. A simple calculation shows that
\begin{eqnarray}
\label{SCF}
\Big\langle U(\mathcal{C}) \, \tilde{W}(\Sigma_{(1|2)}) \Big\rangle = e^{i \alpha \, \tilde{q} \, {\rm SLink}(\mathcal{C}, \Sigma_{(1|2)})} \Big\langle \tilde{W}(\Sigma_{(1|2)}) \Big\rangle  \, ,
\end{eqnarray}
where, once again the charge is given by the super-linking number
\begin{eqnarray}
\label{SCG}
{\rm SLink}(\mathcal{C}, \Sigma_{(1|2)}) = \int_{\SSM}   \mathbb{Y}^{(2|0)}_{\Sigma_{(1|2)}} \wedge \Theta^{(1|2)}_{\Omega_{(2|0)}} = 
 \int_{\SSM}   \Theta^{(1|0)}_{\Omega_{(2|2)}} \wedge  \mathbb{Y}^{(2|2)}_{\mathcal C} \, , 
\end{eqnarray}
with $\mathbb{Y}^{(2|0)}_{\Sigma_{(1|2)}}  = d \Theta^{(1|0)}_{\Omega_{(2|2)}}$ and $\mathbb{Y}^{(2|2)}_{\mathcal C}  = d \Theta^{(1|2)}_{\Omega_{(2|0)}} $.

\section{Conclusions}\label{sec:conclusions}

We presented a geometrical construction of high-form symmetries within the framework of supersymmetric theories. Instead of relying on the conventional approach of superspace, we used the robust framework of supergeometry, which combines the advantages of superspace, differential geometry, and Cartan calculus. 

In our approach, we achieve geometrical integration on supermanifolds by introducing integral forms. This framework enables us to translate well-known results into the realm of supersymmetric theories and supergravity. We provided a comprehensive analysis of charge and charged operators, both in general terms and for several models spanning different dimensions.

Additionally, we constructed new higher-form symmetries using supergeometrical cocycles and performed an extensive analysis of mixed anomalies in this context. Our findings open up possibilities for interesting generalizations to supergravity models and the exploration of symmetries in this field. 

A natural extension of our results involves non-abelian gauge theories and their suitable interactions with matter fields. This will undoubtedly make the construction of charged operators more interesting and challenging. Furthermore, the implications for the physical Hilbert space and the space of observables remain to be explored.

We aim to translate our findings into the construction of non-invertible symmetries within this framework, which we plan to pursue in future work.

Lastly, we have not addressed the case of Voronov-Zorich superforms \(J^{(p|q)}\) with \(q < m\), though these forms could potentially lead to new interesting conserved currents. They may be associated with super-hypersurfaces that have a reduced number of fermionic coordinates. However, the mathematics involved in the construction of these objects is considerably more complex, as the vector spaces required to describe them are naturally infinite-dimensional. Although there have been some new efforts to incorporate them into the discussion, further mathematical tools still need to be developed. 

\section*{Acknowledgements}
We would like to thank E. Torres, I. Valenzuela and G. Tartaglino-Mazzucchelli for useful discussions. PAG thanks CERN TH-Department for hospitality, where most of this work has been done. PAG is partially support by INFN grant {\it Gauge Theory, Supergravity, Strings (GSS 2.0)}. SP is partially supported by the INFN grant {\it Gauge and String Theory (GAST)}. P.A.G. would like to thank the Mittag-Leffler Institute of the Royal Swedish Academy of Science and the organizers of the workshop "Cohomological Aspects of Quantum Field Theory" where this work has been concluded. 

\appendix

\section{Superforms and Integral Forms}\label{app:superforms}

In this appendix, we briefly summarize the main notions of supercalculus in supermanifolds, sticking to the geometric approach introduced in \cite{Witten:2012bg}. We report only what is needed for the comprehension of the present paper, referring to \cite{Witten:2012bg, 
Cremonini:2019aao,
Castellani:2015paa,
Catenacci:2018xsv, 
Catenacci:2019ksa}
for a more complete introduction. 

Given a supermanifold ${\mathcal M}^{(n|m)}$, parametrized by the coordinates $z^A = (x^a, \theta^\alpha)$ where $a=0, \dots, n-1$ and 
$\alpha =1, \dots, m$, we can construct the exterior algebra either  
from the canonical 1-form basis $dx^a, d\theta^\alpha$ or from the anholonomic basis $V^a, \psi^\alpha$ (a.k.a. supervielbeins). The differential is given by 
\begin{eqnarray}
    \label{inA}
    d = dx^a \partial_a + d\theta^\alpha \partial_\alpha  = 
    V^a \partial_a + \psi^\alpha D_\alpha  \, ,
\end{eqnarray}
where $D_\alpha$ is the covariant derivative, while $dx^a$ and $d\theta^\alpha$ are anticommuting and commuting differentials, respectively. 

We define the space of superforms $\Omega^{(\bullet|0)} = \sum_{p=0}^\infty \Omega^{(p|0)}$ where $p$ is the form degree. The second degree denotes the picture number, which is zero for superforms \cite{Belopolsky:1997jz}. Locally, a superform can be decomposed as
\begin{eqnarray}
    \label{inB}
    \omega^{(p|0)} = \sum_{l=0}^p \omega_{[a_1 \dots a_l](\alpha_{l+1} \dots \alpha_{p})}(x,\theta) V^{a_1} \dots V^{a_l} 
    \psi^{\a_{l+1}} \dots \psi^{\a_p} \, ,
\end{eqnarray}
where the components $\omega_{[a_1 \dots a_l](\alpha_{l+1} \dots \alpha_{p})}(x,\theta)$ are superfields. There is no upper bound for the form number $p$ since $\psi$'s are commuting variables w.r.t. 
the wedge product $\psi^\alpha \wedge \psi^\b = \psi^\b 
\wedge \psi^\a$. 

The action of the differential $d$ on a superform increases the form number 
\begin{eqnarray}
    \label{inC}
    d: \Omega^{(p|0)} \longrightarrow \Omega^{(p+1|0)} \, .
\end{eqnarray}

In addition, using the wedge product, the superforms form an algebra 
\begin{eqnarray}
    \label{inD}
    \wedge &:& \Omega^{(p|0)} \times \Omega^{(q|0)} \longrightarrow \Omega^{(p+q|0)} \nonumber \\
    && (\omega^{(p|0)}, \omega^{(q|0)}) \longrightarrow \omega^{(p+q|0)} =  \omega^{(p|0)} \wedge\omega^{(q|0)} \, .
\end{eqnarray}

On the dual side, we introduce the integral forms. To this end,  
we need an additional ingredient 
\begin{eqnarray}
    \Lambda^{(0|m)} = \prod_{\a=1}^m \delta(\psi^\alpha) \, ,
\end{eqnarray}
which is the product of all Dirac delta functions of $\psi$'s. It 
satisfies the following property 
\begin{eqnarray}
    \label{inE}
    \psi^\b \Lambda^{(0|m)}  = 0\,, \qquad \forall \; \b=1, \dots, m \, .
\end{eqnarray}

As in conventional geometry, we can introduce the contraction operator $\iota_X$ along a 
vector field $X$ on the manifold, acting on the exterior algebra. The contraction operator 
$\iota_X$ reduces the form degree by one unit. 
In the present context, we can define the contraction operator along the even-coordinate 
vector field $\frac{\partial}{\partial x^a}$ and the one along the odd-coordinate vector field 
$\frac{\partial}{\partial \theta^\alpha}$. However, it is more convenient to introduce covariant contractions.
Along the covariant vector field $D_\alpha$ we define
\begin{eqnarray}
    \label{inF}
    \iota_\alpha \equiv \iota_{D_\alpha} = \frac{\partial}{\partial \psi^\alpha} \, ,
\end{eqnarray}
whereas along bosonic vectors $\partial_a \equiv \frac{\partial}{\partial x^a}$ we define 
\begin{eqnarray}
    \label{inF}
    \iota_a \equiv \iota_{\partial_a} = \frac{\partial}{\partial V^a} \, ,
\end{eqnarray}
They act on superforms \eqref{inB} and are dual to the supervielbein $(V^a, \psi^\alpha)$, 
\begin{eqnarray}
    \label{inG}
    \iota_\alpha \psi^\b = \delta_\alpha^\beta\,, ~~~~~
    \iota_\alpha V^a = 0 \,, ~~~~
    \iota_a \psi^\alpha =0\,, ~~~~ \iota_a V^b = \delta^b_a \, .
\end{eqnarray}
The commutation relation among  $\iota_\alpha$ and $\iota_a$ are 
\begin{align}
    \iota_a \iota_b = - \iota_b \iota_a\,, ~~~~~
    \iota_\a \iota_\b = \iota_\b \iota_\a\,, ~~~
    \iota_\a \iota_b = \iota_b \iota_\a\, . ~~~
\end{align}

The contraction operator can act as a derivative on  $\Lambda^{(0|m)}$. Therefore, we can build integral forms $\Omega^{(\bullet|m)}$ 
as follows 
\begin{eqnarray}
    \label{inH}
\omega^{(p|m)} = 
\sum_{l=0}^p \omega_{[a_1 \dots a_l](\alpha_{l+1} \dots \alpha_{p})}(x,\theta) V^{a_1} \dots V^{a_l} 
    \iota_{\a_{l+1}} \dots \iota_{\a_{p}}  \Lambda^{(0|m)}  \, . 
\end{eqnarray}
Note that there are no $\psi$'s in this expansion. A $\psi$ either is canceled by \eqref{inE} or reduces the number of $\iota_\a$'s thanks to  the following integration-by-part rule 
\begin{eqnarray}
    \label{inI}
    \psi^\alpha \iota_\b \Lambda^{(0|m)} = - \delta_\a^\b \Lambda^{(0|m)} \, .
\end{eqnarray}
Note that the form number of an integral form goes from $p$ to $-\infty$ since the contractions act, reducing the form number. 
The differential operator $d$ acts on integral forms by increasing the form number, as for superforms 
\begin{eqnarray}
    \label{inL}
    d: \Omega^{(p|m)} \longrightarrow \Omega^{(p+1|m)} \, .
\end{eqnarray}
Integral forms can be multiplied by superforms 
\begin{eqnarray}
    \label{inDA}
    \wedge &:& \Omega^{(p|0)} \times \Omega^{(q|m)} \longrightarrow \Omega^{(p+q|m)} \nonumber \\
    && (\omega^{(p|0)}, \omega^{(q|m)}) \longrightarrow \omega^{(p+q|m)} =  \omega^{(p|0)} \wedge\omega^{(q|m)} \, .
\end{eqnarray}
Instead, multiplying two integral forms leads to a vanishing result due to the properties of the Dirac delta $\delta(\psi^\alpha)$. 

The {\it integral} forms form a differential complex which is isomorphically dual to the complex of superforms.
The notion of the Hodge dual can be established
\begin{eqnarray}
    \label{hodgeA}
    \star: \Omega^{(p|q)} \longrightarrow 
    \Omega^{(p|m-q)}
\end{eqnarray}
and the details are given in 
\cite{Castellani:2015ata,Castellani:2015dis,Castellani:2023tip,Castellani:2016ezd}. In our work, we have considered only two special cases, namely those with $q=0$ and $q=m$. For these two cases we verified the isomorphism, which easily follows from the fact that they are finite dimensional spaces. For the generic case, the isomorphism might require a deeper analysis. In that respect, we also mention the related work \cite{VZ}.

The relevant feature of the integral forms is the they can be integrated on suitable super-cycles (hyper surfaces embedded into a supermanifold). In this way, we can obtain a well-defined expression $$\int_{\mathcal{M}^{(n|m)}} F^{(p|0)} \wedge \mathbb{Y}^{(n-p|m)}_{\Sigma_p}$$ for a topological charge operator as discussed in the text.

\section{The Super-Linking number}\label{app:linking} \setcounter{equation}{0}
 
In this appendix, we review the definition of super-linking numbers in supermanifolds. 
To this end, we first recall briefly 
the definition of \emph{linking number} in bosonic manifolds $\mathcal{M}$, and then we generalize it to supermanifolds.
We will stick to the definition given in terms of Poincar\'e duals to hypersurfaces (see, for instance, \cite{botttu}), as it allows for a straightforward generalization to super-hypersurfaces in supermanifolds. 

In a generic $n$-dimensional manifold ${\cal M}^{(n)}$ we consider two hypersurfaces $\sigma_{p_1}, \sigma_{p_2}$ of dimension $p_1$ and $p_2$ respectively, with embedding equations $\phi_k(x) = 0$, $k=1, \dots , n-p_1$ and $\psi_k(x) = 0$, $k=1, \dots , n-p_2$. The corresponding Poincar\'e dual operators localizing on the two submanifolds are $p_1$- and $p_2$-forms, given by 
\begin{equation}
\label{eq:PCOapp}
\mathbb{Y}_{\sigma_{p_1}}^{(n-p_1)} = \prod_{k=1}^{n-p_1} d \phi_k \delta ( \phi_k ) \equiv d \Theta_{\Omega_{p_1+1}}^{(n-p_1-1)} \; , \qquad \mathbb{Y}_{\sigma_{p_2}}^{(n-p_2)} = \prod_{k=1}^{n-p_2} d \psi_k \delta ( \psi_k ) \equiv d \Theta_{\Omega_{\s_{p_2+1}}}^{(n-p_2-1)} \, ,
\end{equation}	
where $\Omega_{p_1+1}, \Omega_{p_2+1}$ are two hypersurfaces whose boundaries are $\sigma_{p_1}$ and $\sigma_{p_2}$, respectively. Their Poincar\'e duals $\Theta_{\Omega_{p_i+1}}$ are easily found by writing one of the delta functions in \eqref{eq:PCOapp} as the derivative of the Heaviside step function and pulling out the differential. 

The linking number between $\sigma_{p_1}$ and $\sigma_{p_2}$ is equivalently defined as
\begin{equation}\label{app:general}
\hspace{-0.1cm} 	
{\rm Link} \left( \sigma_{p_1} , \sigma_{p_2} \right) = \int_{\mathcal{M}^{(n)}} \mathbb{Y}_{\sigma_{p_1}}^{(n-p_1)}  \wedge \Theta_{\Omega_{\sigma_{p_2+1}}}^{(n-p_2-1)} \, , \quad {\rm or} \, \quad 
{\rm Link}  \left( \sigma_{p_1} , \sigma_{p_2} \right) 	=   \int_{\mathcal{M}^{(n)}}   \Theta_{\Omega_{\sigma_{p_1+1}}}^{(n-p_1-1)} \wedge \mathbb{Y}_{\sigma_{p_2}}^{(n-p_2)} \, .
\end{equation}
This expression is non-vanishing if and only if $p_1 + p_2 = n-1$. Therefore, when assigned the manifold dimension, this constraint selects the dimensions of submanifolds that can intertwine. 

For a three-dimensional manifold, it can be proven that definition \eqref{app:general} applied to two closed (oriented) curves $\gamma_1$ and $\gamma_2$ defined by the two sets of equations $\phi_1(\vec{x}) = \phi_2(\vec{x}) =0$ and $\psi_1(\vec{x}) = \psi_2(\vec{x}) =0$, is equivalent to the Gauss' formula for the linking number of $\gamma_1, \gamma_2$
\begin{equation} \label{Gauss}
	{\rm Link}  \left( \gamma_1 , \gamma_2 \right) = \oint_{\gamma_1} \oint_{\gamma_2} \frac{(\vec{x}_1 - \vec{x}_2)}{|| \vec{x}_1 - \vec{x}_2 ||^3} \,   d \vec{x}_1 \wedge d \vec{x}_2 \, ,
\end{equation}
where $\vec{x}_1$ and $\vec{x}_2 $ are the position vectors on the two loops. For proof of the equivalence of the two prescriptions, we refer to \cite{Cremonini:2020mrk,Cremonini:2020zyn}).

Still, in three dimensions, another interesting example is the evaluation of the linking number between a point $P =
(x_{1,P}, x_{2,P},x_{3,P})$  and a surface $\sigma$, embedded by $\phi(\vec{x}) = 0$. Assigning the corresponding Poincar\'e duals
\begin{eqnarray}
	\nonumber \mathbb{Y}_P^{(3)} &=& dx^1 \delta ( x_1 - x_{1,P} ) dx^2 \delta ( x_2 - x_{2,P} ) dx^3 \delta ( x_3 - x_{3,P} )  \\
	\mathbb{Y}_\sigma^{(1)} &=& d \phi \delta ( \phi ) = d \Theta ( \phi ) \equiv d \Omega_\sigma^{(0)} \, ,
\end{eqnarray}
the linking number is a well-defined three-dimensional integral of a 3-form, whose value is 
\begin{equation} 
	{\rm Link}  \left( P , \sigma \right) = \int_{\mathcal{M}^{(3)}} \mathbb{Y}_{P}^{(3)}  \wedge \Omega_\sigma^{(0)}  
= \int_{\mathcal{M}^{(3)}} d^3x \, \delta^{(3)} ( x - x_P ) \, \Theta ( \phi ) = \begin{cases}
		1 \ , \ \text{if} \; P \in \sigma \\
		0 \ , \ \text{if} \; P \notin \sigma
	\end{cases}
\end{equation}

\vskip 10pt 
We now generalize definition \eqref{app:general} to the case of a $(n|m)$-dimensional supermanifold. Given a purely bosonic $(p_1|0)$-dimensional hypersurface $\Sigma_{(p_1|0)}$ and a $(p_2|m)$-dimensional super-hypersurface $\Gamma_{(p_2|m)}$, the corresponding Poincar\'e duals are\footnote{The definition of super linking number could be generalized to the case of generic pseudo-surfaces of dimensions $(p|q)$ with $0 < q < m$. However, since in the body of the paper we only deal with bosonic surfaces ($q=0)$ and supersurfaces ($q=m$), here we stick only to these two cases. }
\begin{eqnarray}\label{appD:super}
\hspace{-0.5cm} \mathbb{Y}_{\Sigma_{(p_1|0)}}^{(n-p_1|m)} = \mathbb{Y}^{(n-p_1|0)}_{\S_{(p_1|0)}} \prod_{\a=1}^m \theta^\a \delta ( d \theta^\a )  \equiv  
d \Theta_{\Omega_{(p_1+1|0)}}^{(n-p_1-1|m)}\; , \qquad 
\mathbb{Y}_{\S_{(p_2|m)}}^{(n-p_2|0)}  \equiv d \Theta_{\Omega_{(p_2+1|m)}}^{(n-p_2-1|0)} \, ,
\end{eqnarray}
where, as before, the right-hand side is obtained by writing one delta function as the derivative of the step function and pulling out the differential. the $\Theta$s are the Poincar\'e dual of a non-compact 
bosonic hypersurface and a super-hypersurface whose boundaries are $\Sigma_{(p_1|0)}$ and $\Sigma_{(p_2|m)}$, respectively. 

Generalizing the previous construction, the \emph{super-linking number} is defined as \cite{Cremonini:2020zyn}
\begin{equation}\label{appD:super2}
	{\rm SLink} (\Sigma_{(p_1|0)}, \Sigma_{(p_2|m)}) = \int_{\SSM} \mathbb{Y}_{\Sigma_{(p_1|0)}}^{(n-p_1|m)} \wedge \Theta_{\Omega_{(p_2+1|m)}}^{(n-p_2-1|0)} =  \int_{\SSM} 
	\Theta_{\Omega_{(p_1+1|0)}}^{(n-p_1-1|m)} \wedge \mathbb{Y}_{\Sigma_{(p_2|m)}}^{(n-p_2|0)} \, .
	\end{equation}
Again, these integrals are well-defined iff the bosonic dimensions satisfy $p_1+p_2=n-1$. In this case, we do not have any constraint on the fermionic dimensions. Their sum already saturates $m$, as we have chosen from the very beginning to link a bosonic surface (odd dimension zero) with a super-hypersurface (odd dimension $m$). 

Whenever the sum of the fermionic dimensions of the two hypersurfaces does not equal $m$, the super-linking number is zero, even if the bosonic dimensions sum up to $(n-1)$. For instance, two purely bosonic hypersurfaces whose bosonic dimensions satisfy the constraint would have a super-linking number equal to zero anyway. This means they can be somehow unlinked by deforming them along the fermionic directions. 

If, instead, the odd dimensions sum up to $m$, the super-linking number \eqref{appD:super2} is well-defined and generally non-vanishing. In this case, thanks to the particular structure of the Poincar\'e duals, it reduces to the ordinary linking number. In fact, considering for instance the case in \eqref{appD:super}, we find
\begin{equation}
\label{eq:reduction}
	\int_{\SSM} \mathbb{Y}^{(n-p_1|0)}_{\S_{(p_1|0)}} \prod_{\a=1}^m \theta^\a \delta ( d \theta^\a )   \wedge \Theta_{\Omega_{(p_2+1|m)}}^{(n-p_2-1|0)}   = \int_{\mathcal{M}^{(n)} \hookrightarrow \SSM} \mathbb{Y}^{(n-p_1|0)}_{\S_{(p_1|0)}}  \wedge \Theta_{\Omega_{(p_2+1|m)}}^{(n-p_2-1|0)} \, ,
\end{equation}
where ${\mathcal M}^{(n)}$ is the bosonic submanifold of $\SSM$. The general reduction \eqref{eq:reduction} can be applied in all the cases. This is somehow not surprising since the linking number is related to the topological nature of the two hypersurfaces, and the fermionic sector never affects the topology.

\section{Operators charged under (g)CW symmetries in $D=(6|8)$} 
\label{sect:CWcurrents}

In this section, we construct explicitly the operators charged under CW and geometric-CW symmetries for the $D=(6|8)$ super-Maxwell theory. We treat both types of currents on the same foot; therefore, we consider the following list of closed currents 
\begin{align}
    \label{SGcwA}
J^{(2|0)} = F\,, ~~~~~~
J^{(4|0)} = \frac12 F\wedge F\,, ~~~~~~
J^{(5|0)} =  \omega^{(3|0)} \wedge F \,, ~~~~
J^{(5|8)} =  \omega^{(3|8)} \wedge F \, ,
\end{align}
where $\omega^{(3|0)}$ and $\omega^{(3|8)}$ are given in \eqref{sixF} and \eqref{sixG}, respectively. 

According to the general strategy described in the main text, the most convenient way to determine the action of symmetry generators on charged operators is to use a path integral approach where we couple the currents to external backgrounds gauge superfields $B$. Specializing the general expression \eqref{eq:Smin} to the present case, we write
\begin{eqnarray}
\label{eq:6Daction2}
S + S_{min} &=& \int_{\SSM} \left( - \frac12 (F - B^{(2|0)}) \wedge \star (F - B^{(2|0)})  \right. \\
&\,& \qquad \quad \left. + J^{(2|0)}\wedge B^{(4|8)} + J^{(4|0)} \wedge B^{(2|8)} +  
    J^{(5|0)} \wedge B^{(1|8)}
    +  
    J^{(5|8)} \wedge B^{(1|0)}
    \right) \, . \nonumber     
\end{eqnarray}
The currents and the background fields are subject to the following gauge transformations
\begin{eqnarray}
\label{SGcwD}
    && F \rightarrow F + d\Lambda^{(1|0)} 
    \, , \qquad  B^{(2|0)} \rightarrow B^{(2|0)} + d \Lambda^{(1|0)} \, ,  \\
   && B^{(4|8)} \rightarrow B^{(4|8)} + d \Lambda^{(3|8)} - d\Lambda^{(1|0)} \wedge B^{(2|8)}\, , \nonumber \\
   && B^{(2|8)} \rightarrow B^{(2|8)} + d \Lambda^{(1|8)} \, , \nonumber \\
   && B^{(1|8)} \rightarrow B^{(1|8)} + d \Lambda^{(0|8)} \, , \nonumber \\
   && B^{(1|0)} \rightarrow B^{(1|0)} + d \Lambda^{(0|0)} \, , \nonumber 
\end{eqnarray}
which leaves the following field strengths
\begin{eqnarray}
    \label{SGcwDA}
    &&G^{(5|8)} = d B^{(4|8)} + B^{(2|0)} \wedge 
    dB^{(2|8)}\,, ~~~~~
    G^{(3|8)} = d  B^{(2|8)}\,, ~~~~~
    \nonumber \\
    &&G^{(2|8)} = d  B^{(1|8)} \, , ~~~~~~~~~~~~~~~~~~~~~~\qquad
    G^{(2|0)} = d  B^{(1|0)} 
\end{eqnarray}
invariant. Under these gauge transformations, and in the absence of topological singularities that spoil the conservation law $dF=0$,  the action in \eqref{eq:6Daction2} develops an anomalous term. Precisely, we find
\begin{eqnarray}
    \label{SGcwE}
    && \hspace{-0.45cm} S(F-B^{(2|0)}) + S_{min}(B) = S(F'-B^{(2|0)'}) + S_{min}(B') \\
    &\,&\hspace{-0.45cm}
    +\!\! \int_{\SSM} \!\!\! \left( d\Lambda^{(1|0)} \wedge B^{(4|8)} - 
    \frac12 d\Lambda^{(1|0)} \wedge d\Lambda^{(1|0)} \wedge B^{(2|8)} 
    +\omega^{(3|0)}\wedge d\Lambda^{(1|0)} \wedge  B^{(1|8)} +
    \omega^{(3|8)}\wedge \Lambda^{(1|0)} \wedge  B^{(1|0)}  \right) . \nonumber
\end{eqnarray}
As discussed in the main text, this anomaly is responsible for charging Wilson-type operators under the N\"oether symmetry generated by $\star F$. Following the same steps as the lower dimensional cases, we find that the corresponding charged operator is a Wilson loop of type 
\eqref{STSL0} with the charge given by the super linking number, which measures the link between the $\Sigma_{(4|8)}$ super surface supporting the symmetry generator and the contour $\mathcal{C}$ defining the Wilson operator. 

Regarding CW symmetries, we start by considering the ''magnetic'' current $J^{(2|0)}$. Mimicking what we have done in section \ref{sec:4D} for the four-dimensional case, we study the action of the generator 
$e^{i\a_m Q_m(\Sigma_{(2|0)})}$, with $Q_m(\Sigma_{(2|0)}) = \int_{\SSM} J^{(2|0)} \wedge \mathbb{Y}_{\Sigma_{(2|0)}}^{(4|8)}$, on supermonopoles described by singular supersurfaces $\Sigma_{(3|8)}$ that spoil the Bianchi identity according to
\begin{equation}
\label{eq:monopoleA}
    dF = q_m \mathbb{Y}_{\Sigma_{(3|8)}}^{(3|0)} \, .
\end{equation}
Writing $\mathbb{Y}_{\Sigma_{(3|8)}}^{(3|0)} = d \Theta^{(2|0)}_{\Omega_{(4|8)}}$, the inclusion of the defect inside the path integral corresponds to turning on $B^{(2|0)}_{\mathcal{D}} = q_m \Theta^{(2|0)}_{\Omega_{(4|8)}}$. At the same time, the insertion of the symmetry generator is equivalent to $B^{(4|8)} = \a_m \mathbb{Y}_{\Sigma_{(2|0)}}^{(4|8)}$. We can get rid of the generator by computing the two-point function between the generator and the defect by performing a $\Lambda^{(3|8)}$-gauge transformation. However, since in the presence of the defect $dF$ is no longer zero everywhere, this transformation generates an anomalous term, which eventually gives rise to the charge of the supermonopole under $Q_m$. In fact, an explicit calculation leads to
\begin{equation}
\label{STSO}
 \Big\langle e^{i \alpha_m Q_m(\Sigma_{(2|0)})}  \mathcal{D}({\Sigma_{(3|8)}}) \Big\rangle = e^{i \alpha_m  \, q_m {\rm SLink}(\Sigma_{(2,0)}, {\Sigma_{(3|8)}})}  \Big\langle \mathcal{D}({\Sigma_{(3|8)}}) \Big\rangle    \, ,
\end{equation}
where the charge is the super-linking number between the two supersurfaces (we set $ \mathbb{Y}^{(4|8)}_{\Sigma_{(2|0)}}=  d\Theta^{(3|8)}_{\Omega_{(3|0)}}$)
\begin{eqnarray}
\label{STSP}
{\rm SLink}(\Sigma_{(2,0)}, \Sigma_{(3|8)}) = 
\int_{\SSM} \Theta^{(2|0)}_{\Omega_{(4|8)}} \wedge \mathbb{Y}^{(4|8)}_{\Sigma_{(2|0)}} = \int_{\SSM} \mathbb{Y}^{(3|0)}_{\Sigma_{(3|8)}} \wedge \Theta^{(3|8)}_{\Omega_{(3|0)}} \, . 
\end{eqnarray}
An alternative but equivalent derivation of \eqref{STSO} could be obtained by using the dual description of the super-Maxwell theory given in terms of the dual gauge field $\tilde A^{(3|8)}$, defined as  $\star F = d \tilde A^{(3|8)}$. 

\vskip 10pt
The magnetic supermonopole introduced above turns out to be also charged with respect to the gCW symmetries. To prove this statement and find the corresponding charge, we start  considering the symmetry associated with the gCW current $J^{(5|0)} = \omega^{(3|0)} \wedge F$ and generated by
\begin{eqnarray}
    \label{COtA}
    U(\Sigma_{(5|0)},\a) = e^{i \alpha \int_{\SSM} \omega^{(3|0)} \wedge F \wedge \mathbb{Y}^{(1|8)}_{\Sigma_{(5|0)}}} \, .
\end{eqnarray}
Its insertion inside the path integral 
corresponds to turning on a background field $B^{(1|8)} = \alpha \mathbb{Y}^{(1|8)}_{\Sigma_{(5|0)}}$ in action \eqref{eq:6Daction2}. 

In the presence of the supermonopole $\mathcal{D}(\Sigma_{(3|8)})$, the $J^{(5|0)}$ current is no longer conserved everywhere. In fact, as a consequence of \eqref{eq:monopole} we have
\begin{equation}
\label{eq:monopole2}
    dJ^{(5|0)} = q_m \, \omega^{(3|0)} \wedge \mathbb{Y}^{(3|0)}_{\Sigma_{(3|8)}} = d \left( q_m \, \omega^{(3|0)} \wedge \Theta^{(2|0)}_{\Omega_{(4|8)}}\right) \, .
\end{equation}

We then evaluate the two-point function $\langle U(\Sigma_{(5|0)}) \mathcal{D}(\Sigma_{(3|8)}) \rangle$, taking into account that the insertion of the $\mathcal{D}$ defect corresponds to turning on $B^{(2|0)}_{\mathcal{D}} = q_m \Theta^{(2|0)}_{\Omega_{(4|8)}}$. Performing a $\Lambda^{(0|8)}$-gauge transformation in \eqref{eq:6Daction2} to get rid of the $U$ generator, we are left with an anomalous term due to the lack of conservation \eqref{eq:monopole2}. This gives rise to an extra phase that can be interpreted as giving the $\mathcal{D}$ charge under the action of $U(\Sigma_{(5|0)})$. Precisely, we find
\begin{equation}
\label{eq:CWcharge}
\Big\langle U(\Sigma_{(5|0)}, \a) \mathcal{D}(\Sigma_{(3|8)}) \Big\rangle    = e^{i\alpha \, q_m \int_{\SSM} \Theta^{(2|0)}_{\Omega_{(4|8)}} \wedge \, \omega^{(3|0)} \wedge Y^{(1|8)}_{\Sigma_{(5|0)}}}
\, \Big\langle  \mathcal{D}(\Sigma_{(3|8)}) \Big\rangle \, . 
\end{equation}

The charge in the exponent is given in terms of an expression that is generally non-trivial and can still be interpreted as a super-linking number. If non-vanishing, this measures the link between the defect supersurface $\Sigma_{(3|8)}$ and a $(2|0)$-supersurface obtained by a three-dimensional foliation of $\Sigma_{(5|0)}$ described by the Poincar\'e dual $\omega^{(3|0)} \wedge Y^{(1|8)}_{\Sigma_{(5|0)}}$.

In order to show that the charge can be non-zero, we  consider taking
\begin{eqnarray}
    \label{COtF}
     \mathbb{Y}^{(1|8)}_{\Sigma_{(5|0)}} = V^{\alpha\alpha'} 
     V^{\beta\beta'} V^{\gamma\gamma'} \epsilon_{\alpha\beta\gamma\rho} 
     \epsilon_{\alpha'\beta'\gamma'\rho'} \epsilon^{AB}\iota^{\rho}_A \iota^{\rho'}_B \delta^{(8)}(\psi)  \, , 
\end{eqnarray}
which is a closed but not exact $(1|8)$-integral form, thus interpretable as the Poincar\'e dual of a $(5|0)$ supersurface. Inserting this expression into the phase of \eqref{eq:CWcharge} and taking into account the definition of $\omega^{(3|0)}$ in \eqref{sixF}, we obtain 
\begin{eqnarray}
    \label{COtFa}
    &&\alpha \, q_m \int_{\SSM} \Theta^{(2|0)}_{\Omega_{(4|8)}} \wedge V^{[\alpha\beta]} \psi^{A}_\alpha 
\psi^{B}_\beta \epsilon_{AB} \wedge 
    V^{\alpha\alpha'} 
     V^{\beta\beta'} V^{\gamma\gamma'} \epsilon_{\alpha\beta\gamma\rho} 
     \epsilon_{\alpha'\beta'\gamma'\rho'} \epsilon^{A'B'}\iota^{\rho}_{A'} \iota^{\rho'}_{B'} \delta^8(\psi) 
     \nonumber \\
    && =
    \alpha \, q_m \int_{\SSM} V^{[\rho\rho']}
    V^{\alpha\alpha'} 
     V^{\beta\beta'} V^{\gamma\gamma'} \epsilon_{\alpha\beta\gamma\rho} 
     \epsilon_{\alpha'\beta'\gamma'\rho'}  \delta^8(\psi) 
    \wedge \Theta^{(2|0)}_{\Omega_{(4|8)}} \, .
\end{eqnarray}
Expanding $ \Theta^{(2|0)}_{\Omega_{(4|8)}} $, in general this results in a non-trivial expression. 

A similar calculation proves that the supermonopole is also charged under $J^{(5|8)}$ in \eqref{SGcwA}.  
In this case, the insertion of the symmetry generator 
\begin{eqnarray}
    \label{COtG}
    U(\Sigma_{(5|8)}, \b) = e^{i \beta \int \omega^{(3|8)} \wedge F \wedge \mathbb{Y}^{(1|0)}_{\Sigma_{(5|8)}}}
\end{eqnarray}
in the path integral corresponds to turning on the background field $B^{(1|0)} = 
\b \mathbb{Y}^{(1|0)}_{\Sigma_{(5|8)}}$. Again,  acting with $U$ on the defect and computing the path integral as done above, we find that the supermonopole carries $J^{(5|8)}$ charge given by
\begin{eqnarray}
    \label{COtH}
    \beta \, q_m \int 
    \omega^{(3|8)} \wedge \mathbb{Y}^{(1|0)}_{\Sigma_{(5|8)}} \wedge \Theta^{(2|0)}_{\Omega_{(4|8)}}
     \, .
\end{eqnarray}

To evaluate explicitly this expression we assume that $\Sigma_{(5|8)}$ is 
described by an algebraic equation $\Phi(x, 
\theta)=0$, whereas $\Omega_{(4|8)}$ is featured  by three algebraic equations 
$\Xi^I(x, \theta)=0$ with $I=1,2,3$. We then define the product 
\begin{eqnarray}
    \label{COtI}
   \mathbb{Y}^{(1|0)}_{\Sigma_{(5|8)}} \wedge \Theta^{(2|0)}_{\Omega_{(4|8)}}  = \delta\left( \Phi\right) d\Phi  \bigwedge_{I=1}^2\delta\left( \Xi^I\right) d\Xi^I
   \,  \Theta\Big(\Xi^3\Big) \, .
\end{eqnarray}
Inserting this expression in \eqref{COtH} together with the explicit form \eqref{sixG} of $\omega^{(3|8)}$, we find
\begin{eqnarray}
    \label{COtJ}
    &&\int_{\SSM}(V^5)_{\alpha\beta} \, \epsilon^{AB}
    \iota^\alpha_A \iota^\beta_B \Big[
    d\Phi\wedge 
    d\Xi^1 \wedge 
    d\Xi^2
    \Big] \delta^{(8)}(\psi) \delta\left( \Phi\right) 
    \prod_{I=1,2}\delta\left( \Xi^I\right) 
    \Theta\Big(\Xi^3 \Big) 
\nonumber \\
&=&     
\int_{\SSM} d^6x d^8 \theta \, \epsilon^{AB}
    \Big[
    \partial_{\alpha\beta}\Phi \wedge 
    D^\alpha_A\Xi^1\wedge 
    D^\beta_B\Xi^2
    \Big] \delta(\Phi) 
    \prod_{I=1,2}\delta\left( \Xi^I\right) 
    \Theta\Big(\Xi^3 \Big) \, , 
\end{eqnarray}
where we have used $V^6 \delta^{(8)}(\psi) = d^6x d^8 \theta$ and the two fermionic superfields inside of the square bracket come from integrating by parts the $\iota$ contractions inside $\omega^{(3|8)}$. 
The remaining delta functions project the coordinates onto the algebraic curve. Since the arguments of the delta's are functions of bosonic and spinorial coordinates, it is convenient to expand them in powers of $\theta$s and first perform the Berezin integral. The final result will generally be a linear combination of derivative terms that do not vanish. 

\vskip 10pt

Finally, we discuss operators charged under the action of the CW generator
\begin{equation}
    U(\Sigma_{(4|0)} ,\a_2) = e^{i \a_2 \int_\SSM (\frac12 F \wedge F) \wedge \mathbb{Y}^{(2|8)}_{\Sigma_{(4|0)}}} \, ,
\end{equation}
which gives rise to a $(1|8)$-higher-integral form symmetry. It is sourced in the path integral by turning on a background gauge field $B^{(2|8)}$ in the action \eqref{eq:6Daction2}. 

The corresponding charged objects are non-local operators supported on $(1|8)$-dimensional supersurfaces in $\SSM$. 
As in the bosonic case \cite{Nakajima:2022feg} (see also the summary in section \ref{sec:maxwell}), the support can be realized as the intersection of two supermonopole surfaces $\Sigma_{(3|8)}$ and $\Sigma_{(3|8)}'$ given above, which have in common one bosonic direction and all the fermionic ones. The insertion of this operator inside a correlation function corresponds to turning on a background field $B_{\mathcal{D}}^{(2|0)} \wedge B_{\mathcal{D}}^{(2|0)'}$ in the path integral, with $B_{\mathcal{D}}^{(2|0)} = q_m \Theta^{(2|0)}_{\Omega_{(4|8)}}$,  $B_{\mathcal{D}}^{(2|0)'}= q_m' \Theta^{(2|0)}_{\Omega_{(4|8)}'}$ and $\partial \Omega_{(4|8)} \cap \partial \Omega_{(4|8)}' = \Sigma_{(1|8)}$. The charge of this $(1|8)$-dimensional defect is then $q_2 = q_m q_m'$. 

In the presence of such a defect, the $J^{(4|0)}$ current is no longer closed, rather 
\begin{equation}
d (\frac12 F \wedge F ) = q_2 \mathbb{Y}_{\Sigma_{(1|8)}}^{(5|0)} \, .
  \end{equation}
Therefore, applying the same procedure adopted for the other symmetries, due to an anomaly arising under a $\Lambda^{(1|8)}$ gauge transformation \eqref{SGcwD}, which removes the symmetry generator, we eventually find 
\begin{equation}
\Big\langle U(\Sigma_{(4|0)}, \a_2) \mathcal{D}(\Sigma_{(1|8)}) \Big\rangle    = e^{i\alpha_2 \, q_2 \int_{\SSM} \Theta^{(1|8)}_{\Omega_{(5|0)}} \wedge \, Y^{(5|0)}_{\Sigma_{(1|8)}}}
\, \Big\langle  \mathcal{D}(\Sigma_{(1|8)}) \Big\rangle \, , 
\end{equation}
where $\partial \Omega_{(5|0)} = \Sigma_{(4|0)}$. Once again, the charge of the operator with respect to the $J^{(4|0)}$-higher-superform symmetry is given by the super-linking number measuring the link between the supersurface supporting the symmetry generator and the one supporting the defect.

\bibliographystyle{JHEP}
\bibliography{super.bib}
\end{document}